\documentclass[useAMS,usenatbib,usegraphicx]{mn2e}
\usepackage{times}
\usepackage{color}
\usepackage{amssymb,amsmath}

\def\la{\raise.5ex\hbox{$&lt;$}\kern-.8em\lower 1mm\hbox{$\sim$}}
\def\ma{\raise.5ex\hbox{$&gt;$}\kern-.8em\lower 1mm\hbox{$\sim$}}

\def\kms{$\rm km\, s^{-1}$}
\def\cm3{$\rm cm^{-3}$}

\def\n0{$\rm n_{0}$}
\def\B0{$\rm B_{0}$}

\def\L12{L$_{12\mu m}$~}
\def\F12{F$_{12\mu m}$~}

\def\fe2{[Fe\,{\sc ii}]}
\def\h2{H$_{2}$}
\def\pp{$\pm$}
\def\mc{$\mu$m}

\title[Molecular Hydrogen and \lbrack Fe\,{\sc ii}\rbrack\ in AGN]{Molecular Hydrogen and \lbrack Fe\,{\sc ii}\rbrack\ in Active Galactic Nuclei III: LINERS and Star Forming Galaxies}

\author[Riffel et al.]{R. Riffel$^1$\thanks{E-mail: riffel@ufrgs.br. Visiting Astronomer at the Infrared Telescope Facility, which is operated by the University of Hawaii
    under Cooperative Agreement no. NCC 5-538 with the National Aeronautics and Space Administration, Office of Space Science, Planetary Astronomy Program.}, 
A. Rodr\'{\i}guez-Ardila$^2$, I. Aleman$^{1}$, M. S. Brotherton$^3$, 
  \newauthor M. G. Pastoriza$^{1}$, C. Bonatto$^{1}$, O. L. Dors Jr.$^4$
 \\
$^{1}$Departamento de Astronomia, Universidade Federal do Rio Grande do Sul - Av. Bento Gon\c calves 9500, Porto Alegre, RS, Brasil. \\
$^2$ Laborat\'{o}rio Nacional de Astrof\'{i}sica/MCT - Rua dos Estados Unidos 154, Bairro das Nac\~oes. CEP 37504-364, Itajub\'{a}, MG, Brasil\\
$^3$ Department of Physics and Astronomy, University of Wyoming, Laramie, WY 82071, USA\\
$^4$ Universidade do Vale do Para\'{i}ba - Av. Shishima Hifumi 2911, Cep 12244-000, S\~ao Jos\'e dos Campos, SP, Brasil
}

\begin{document}

\date{}


\maketitle

\label{firstpage}

\begin{abstract}

We study the kinematics and excitation mechanisms of \h2\ and \fe2\ lines in a sample of 67 emission-line galaxies with Infrared Telescope Facility SpeX near-infrared (NIR, 0.8-2.4\mc) spectroscopy together with new photoionisation models. \h2\ emission lines are systematically narrower than narrow-line region (NLR) lines, suggesting that the two are, very likely, kinematically disconnected. The new models and emission-line ratios show that the thermal excitation plays an important role not only in active galactic nuclei but also in star forming galaxies. The importance of the thermal excitation in star forming galaxies may be associated with the presence of supernova remnants close to the region emitting \h2\ lines. This hypothesis is further supported by the similarity between the vibrational and rotational temperatures of \h2. 
We confirm that the diagram involving the line ratios \h2\ 2.121\mc/Br$\gamma$ and \fe2\ 1.257\mc/Pa$\beta$ is an efficient tool for separating emission-line objects according to their dominant type of activity. We suggest new limits to the line ratios in order to discriminate between the different types of nuclear activity.

\end{abstract}


\section{Introduction}

Molecular gas emission has been detected in the inner few tens of parsecs 
of the central region of emission-line galaxies \citep[e.g.][]{ag02,gra03,rogerio-atlas06,rogemar_eso06,rogemar_4051_08,rogemar_4151_10,rogemar_1066_11,rogemar_1157_11}. However, the origin of these lines remains unclear.  They appear in objects classified from starburst to 
active galactic nucleus-dominated. In active galactic nuclei (AGNs), unification models
predict the presence of a dusty molecular torus that obscures the central source from some viewing angles, thus creating apparently different AGN classes
(e.g. Seyfert 1 and Seyfert 2 galaxies). In order to block the radiation coming from the central source, the torus must have a very large optical depth (tens of magnitudes), thus serving as 
a natural reservoir of molecular gas. For example, \citet[][ see their Fig.~4]{gra03} propose that the torus is the source of the \h2\ emission-lines observed in NGC\,1068. However, Integral Field Unit (IFU) spectroscopy of this source shows that the molecular gas is distributed along the galaxy and is spatially correlated with a young stellar population \citep[30 Myr old,][]{thaisa12}.

Observational evidence has confirmed that circumnuclear/nuclear starbursts can coexist in objects harbouring AGNs \citep[e.g.][]{miz94,imdul00,stu02,con02,ara03,imwa04,thaisa05,shi06}. However, in the case of low-luminosity AGNs, no strong starburst evidence is found, but significant fractions of intermediate-age stars are detected \citep[][ and references therein]{iva00,cid04,rosa04,cid05,cid10}. Thus, if star formation is taking place or occurred recently,  the presence of molecular clouds associated with the star forming regions should leave spectral signatures that may contaminate the
AGN spectrum. Therefore, whether the H$_{2}$ lines observed in AGNs are directly associated with either the molecular dusty torus or the circumnuclear
gas remains an open question. Alternatively, the molecular gas may be directly associated with the AGN, but distributed within the circumnuclear region. This later scenario is supported by the fact that the \h2\ gas is most probably arranged in a disc surrounding the nucleus \citep{rka02,rka03,neum07}. In this picture, 
shocks by radio jets and X-ray heating are plausible excitation mechanisms for the \h2\ molecule \citep{knop96,rka02,oli12}.

As shown by \citet{ara04, ara05} and \citet{rogerio-atlas06} in the study of 51 objects, mostly AGNs, molecular hydrogen is detected in almost all Seyfert~2 (Sy~2) and in about 80\% of the Seyfert~1 (Sy~1) sources. Interestingly, the four starburst galaxies in \citet{rogerio-atlas06} also show the presence of molecular hydrogen emission. However, these previous works have not included a statistically significant number of Low-ionisation nuclear emission-line region (LINER) and Star Forming Galaxies (SFGs). In fact, up to now, only a few investigations have included these type of sources \citep[e.g.][]{lar98}. These classes deserve separate analyses, as it is expected that the active nucleus has very little  influence on their emission-line spectrum. 

The \h2\ infrared emission lines may be produced by the radiative de-excitation following collisional excitation (thermal process) or UV (6-18 eV) photon absorption (non-thermal process). \h2\ may also be formed in an excited state, after which de-excitation also leads to emission of IR lines. This mechanism is, however, not very important in a hot or strongly UV/X-ray irradiated gas \citep[e.g.][]{Aleman_Gruenwald_2011} as AGNs. \h2\ IR emission can be significant in regions illuminated by UV photons \citep{bvd, sd89} or X-rays \citep{mht96,lm83}, as well as in shocks \citep{hm89}. In each case, a different emission-line spectrum is expected and, therefore, the relative intensities among the \h2\ emission-lines may be used to discriminate between the powering mechanisms \citep[e.g.][]{mo94}, keeping in mind that multiple emitting regions and multiple mechanisms may simultaneously be present \citep{ara04,ara05,rogemar_eso06,rogemar_4051_08,rogemar_4151_09,rogemar_1157_11}.

\citet{ara04,ara05}  carried out a pioneering study of the \h2\ excitation mechanisms in a sample of Seyfert galaxies. 
The innovation of these investigations compared to previous ones \citep{vgh97,lar98,rka02,rka03} is that their
integrated spectra cover the inner 300~pc in most of the sample, minimizing the host galaxy contamination and maximizing the number of diagnostic lines observed.  In addition,  \fe2\ and \h2\ emission-lines are usually suggested to form in the same region, and the evidence accumulated shows that 
forbidden iron emission in AGNs can have different sources, but very likely, all directly related to the central engine \citep[][ and references therein]{fw93,gvh94,sfbw96,alh97,mkt00,ara04, ara05,davies05,rogemar_eso06,rogemar_4051_08,rogemar_4151_09,rogemar_1157_11,oli12}. Thus, it is interesting to study the excitation mechanisms leading to \h2, together with the processes responsible for the \fe2\ emission-lines.

With the above in mind, we present a study of the excitation mechanisms leading
to the emission of \h2\ and \fe2\ for a sample of nearby LINERs and SFGs.  
We present 16 new spectra acquired with the same instrumentation employed by \citet{ara04,ara05} and \citet{rogerio-atlas06}, with simultaneous observations of the
$JHK$-bands, eliminating differences in aperture and seeing across the bands. The samples combined correspond to  an homogeneous data set with $\sim$ 70 sources. To date this is the largest number of galaxies used to study the nuclear excitation mechanisms of \h2\ and \fe2.

This paper is structured as follows. \S~\ref{obs} presents the observations 
and data reduction steps. \S~\ref{kin} discusses the kinematics of the \h2\ and \fe2\ gas inferred from their line profiles. In \S~\ref{exit}, the excitation mechanisms of the \h2\ IR lines in AGN are discussed using new photoionisation models for AGNs and Starbusts. The calculated emission is compared to the observed \h2\ line ratios in a 1-0~S(2)/1-0~S(0) vs 2-1~S(1)/1-0~S(1) diagnostic diagram.
\S~\ref{diagnostic_diagrams} analyses the line ratios H$_2$ 2.121$\mu m$/Br$\gamma$ and \fe2\ 1.257$\mu m$/Pa$\beta$ for different emission-line objects and the apparent correlation between these ratios over a large range of values.  It also discusses the difference in these ratios for different nuclear activity type. Concluding remarks are given in \S~\ref{fin}. 
A Hubble constant of 75~km~s$^{-1}$~Mpc$^{-1}$ will be used throughout this work in order to be consistent with our previous studies.

\section{Observations and data reduction} \label{obs}

Near-infrared (NIR) spectra in the range 0.8-2.4 $\mu$m were obtained on October 4,6, and 7 in 2010 with the
SpeX spectrometer \citep{ray03} attached to the NASA 3\,m Infrared Telescope Facility (IRTF). 
The detector is a 1024$\times$1024 ALADDIN 3 InSb array with a spatial scale of
0.15$''$/pixel. Simultaneous wavelength coverage was obtained by means of prism cross-dispersers.  A 0.8$'' \times$15$''$ slit was used during the
observations, giving a spectral resolution of 360~\kms. Both the the arc lamp spectra and the night sky spectra are consistent with this value.
The seeing varied between 0.4$''$--0.7$''$ over the different nights.

Due to the extended nature of the sources, the observations were done by nodding in an Object-Sky-Object pattern with typical individual integration times of 120\,s
and total on-source integration times between 18 and 58 minutes. During the observations, A0\,V stars were observed near each target to provide telluric
standards at similar air masses. They were also used to flux calibrate the sample.  Table~\ref{log} shows the observation log. The galaxies are listed in
order of right ascension, and the number of exposures refers to on source integrations.

Spectral extraction and wavelength calibration were
performed using the {\sc spextool}, a software developed and provided
by the SpeX team for the IRTF community \citep{cvr03}\footnote{{\sc spextool} is available from the
IRTF web site at http://irtf.ifa.hawaii.edu/Facility/spex/spex.html},
following the same procedures as described by \citet{ara04}. Column~10 of Table~\ref{log}  
lists the radius of the integrated region, with centre at the peak of the continuum light distribution for every object of the sample. 
No effort was made to extract spectra at positions different from the nuclear region, even though some objects show evidence of extended emission. 
Telluric absorption correction and flux calibration were applied to the individual 1-D spectra by means of the IDL 
routine {\it xtellcor} \citep{vcr03}.

\begin{figure*}
\includegraphics[width=\textwidth]{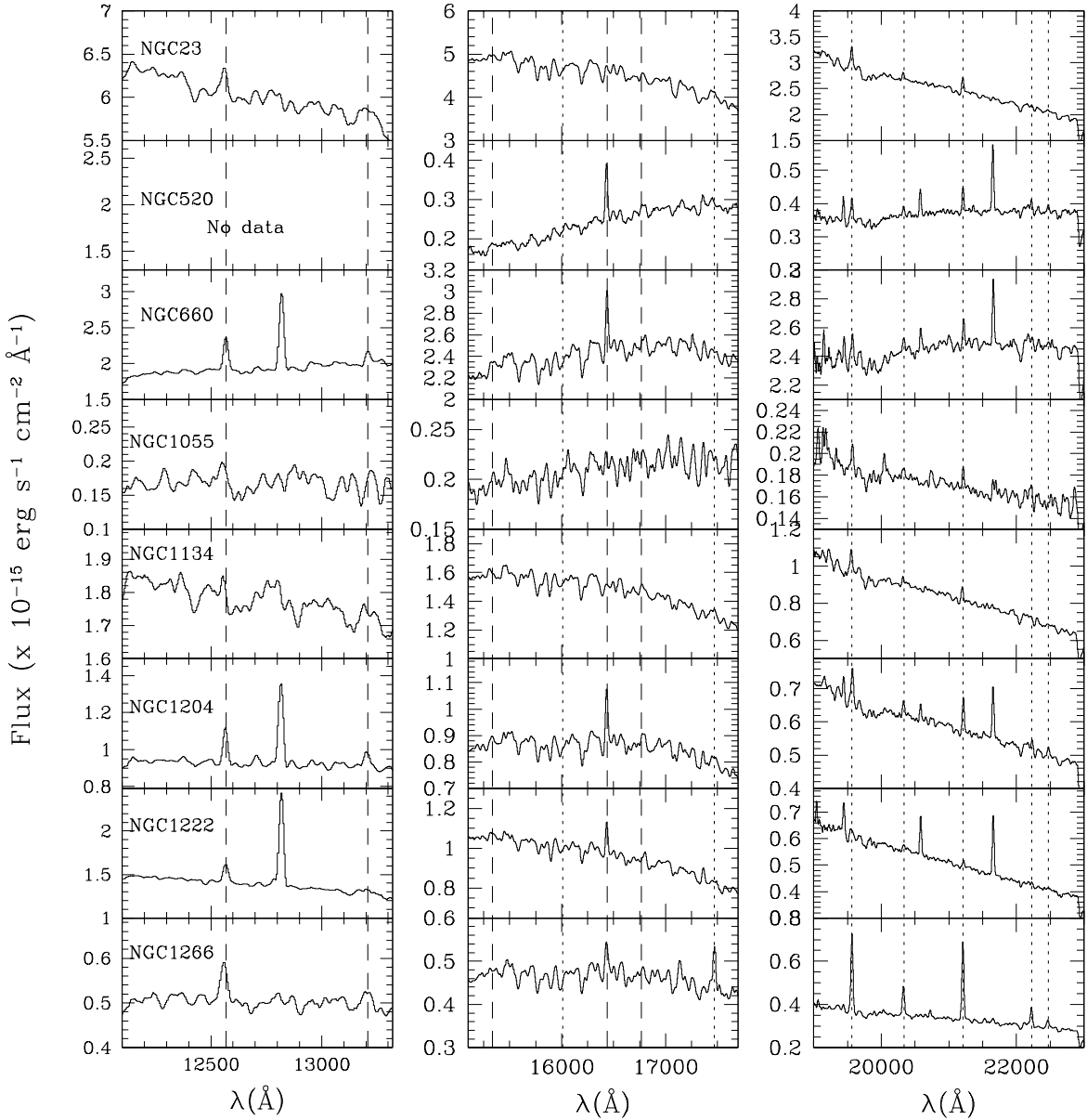} 
\caption{SpeX final reduced spectra, in the Earth's frame of reference, centred
near Pa$\beta$ (left panel), the H-band (1.60$\mu$m, middle panel),
and Br$\gamma$ (right panel). The observed flux is in
units of 10$^{-15}$ erg cm$^{-2}$ s$^{-1}$ \AA$^{-1}$. The identified
\fe2\ (dashed lines) and \h2\ lines (dotted lines) are marked in the
spectra.}
\label{plot1}
\end{figure*}

\begin{figure*}
\includegraphics[width=\textwidth]{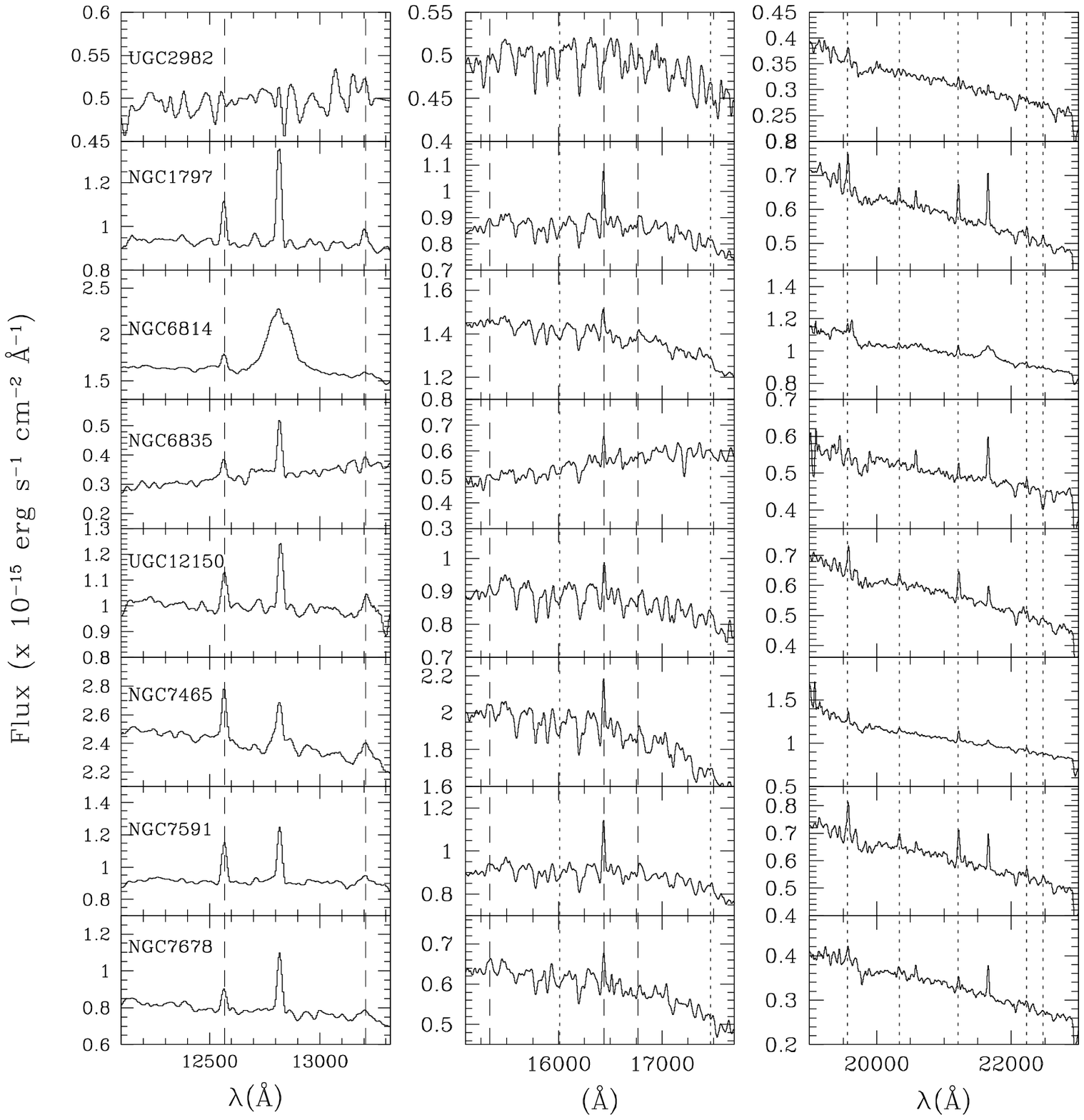} 
\caption{Same as Fig.~\ref{plot1} for the remaining objects.}
\label{plot2}
\end{figure*}

The 1-D wavelength and flux calibrated spectra were then corrected for redshift, determined from the average $z$ measured from the position of [S\,{\sc iii}]
0.953$\mu$m, Pa$\delta$, He\,{\sc i}~1.083$\mu$m, Pa$\beta$ and Br$\gamma$.  Final reduced spectra, in the spectral regions of interest in this work, are plotted in Figures~\ref{plot1} and \ref{plot2}.

\begin{table*}
\centering
\caption{Observation log and basic properties of the sample.\label{log}} 
\begin{tabular}{lclccccclcc}
\hline \hline
\noalign{\smallskip}
Source   & Activity    &   Activity    &   Obs.  Date	&     Exp. Time      &  Airmass  &   RA 	 &     DEC	&  ~~~~~~~~z   & Pos. Angle & radius$^b$    \\ 
         &             &   Reference   &		&      (s)	     &       &  	     &  	    &		   & (deg)  &  (pc)    \\
\noalign{\smallskip}
\hline
NGC 23    & SFG        &  1	         & 2010 10 07	&    29 $\times$ 120 &  1.04 &  00h09m53.4s  &   +25d55m26s & 0.0157202    & 330 &  609    \\ 
NGC 520   & SFG        &  2	         & 2010 10 04	&    16 $\times$ 120 &  1.04 &  01h24m35.1s  &   +03d47m33s & 0.0080367    & 300 &  311    \\ 
NGC 660   & LINER/HII  &  2,3,4          & 2010 10 06	&    24 $\times$ 120 &  1.01 &  01h43m02.4s  &   +13d38m42s & 0.0029152    & 33  &  107    \\ 
NGC 1055  & LINER/HII  &  2,3,4          & 2010 10 04	&    16 $\times$ 120 &  1.07 &  02h41m45.2s  &   +00d26m35s & 0.0036267    & 285 &  210    \\ 
NGC 1134  & SFG        &  5	         & 2010 10 04	&    16 $\times$ 120 &  1.11 &  02h53m41.3s  &   +13d00m51s & 0.0129803    & 0   &  503    \\ 
NGC 1204  & LINER      &  6 	         & 2010 10 07	&    16 $\times$ 120 &  1.23 &  03h04m39.9s  &   -12d20m29s & 0.0154058    & 66  &  597    \\ 
NGC 1222  & SFG        &  7	         & 2010 10 06	&    24 $\times$ 120 &  1.13 &  03h08m56.7s  &   -02d57m19s & 0.0082097    & 315 &  270    \\ 
NGC 1266  & LINER      &  7	         & 2010 10 07	&    18 $\times$ 120 &  1.09 &  03h16m00.7s  &   -02d25m38s & 0.0077032    & 0   &  298    \\ 
UGC 2982  & SFG        &  8	         & 2010 10 04	&    9  $\times$ 120 &  1.11 &  04h12m22.4s  &   +05d32m51s & 0.0177955    & 295 &  689    \\ 
NGC 1797  & SFG        &  1	         & 2010 10 07	&    16 $\times$ 120 &  1.23 &  05h07m44.9s  &   -08d01m09s & 0.0154111    & 66  &  597    \\ 
NGC 6814  & Sy~1       &  7	         & 2010 10 07	&    16 $\times$ 120 &  1.17 &  19h42m40.6s  &   -10d19m25s & 0.0056730    & 0   &  220    \\ 
NGC 6835  & ?	       &  -	         & 2010 10 06	&    22 $\times$ 120 &  1.21 &  19h54m32.9s  &   -12d34m03s & 0.0057248    & 70  &  166    \\ 
UGC 12150 & LINER/HII  &  9	         & 2010 10 04	&    15 $\times$ 120 &  1.08 &  22h41m12.2s  &   +34d14m57s & 0.0214590    & 37  &  748    \\ 
NGC 7465  & LINER/Sy~2 & 10	         & 2010 10 06	&    12 $\times$ 120 &  1.03 &  23h02m01.0s  &   +15d57m53s & 0.0066328    & 340 &  257    \\ 
NGC 7591  & LINER      &  7	         & 2010 10 07	&    16 $\times$ 120 &  1.03 &  23h18m16.3s  &   +06d35m09s & 0.0165841    & 0   &  642    \\ 
NGC 7678  & SFG        &  11	         & 2010 10 04	&    16 $\times$ 120 &  1.01 &  23h28m27.9s  &   +22d25m16s & 0.0120136    & 90  &  419    \\ 
\noalign{\smallskip}
\hline
\end{tabular}
\begin{list}{Table Notes:}
\item SFG: Star Forming Galaxies (Starburst or H{\sc ii} galaxies). LINER/HII were assumed as pure LINERs in the text. References - 1: \citet{balzano83}; 2: \citet{ho97b}; 3: \citet{ho97}; 4: \citet{filho04}; 5: \citet{condon02}; 6: \citet{stu06}; 7: \citet{parreira10}; 8: \citet{schmitt06}; 9: \citet{veilleux95}; 10: \citet{ferruit00}; 11: \citet{goncalves98}.
\end{list}
\end{table*}

\section{on the molecular gas kinematics}\label{kin}

The main goal of this section is to discuss how the velocity width of the H$_{2}$\,2.1213$\mu$m emission-line
compares to that of \fe2 and [S{\sc iii}] in different objects. This provides important constraints to the location of the molecular gas, and allows a comparison between this region with the different types of nuclear activity. It is worth mentioning that the nuclear activity used here (Table~\ref{log}) comes primarily from optical studies, and was taken from the NASA Extragalactic Database (NED). However, NIR emission-line ratios may reveal the presence of ``hidden" AGNs (see Sec.~\ref{diagnostic_diagrams}).

Previous results, mainly on Seyfert galaxies, are controversial.  An imaging study of \h2\ emission \citep{q99} shows that the H$_{2}$ is distributed on scales of a few hundred parsecs from the nucleus. In addition, the fact that the \h2\ emission was found to coincide with [O\,{\sc iii}] and H$\alpha$+[N\,{\sc ii}] for some sources led the authors to suggest that the molecular gas may follow the narrow line region (NLR) gas distribution. \citet{schi99} suggest that the molecular gas in NGC~1068 and NGC~3727 originates from a warped disc with a radius smaller than $\sim$ 75~pc. These latter results are supported by those of \citet{ara04,ara05}, who found that the H$_{2}$~2.1213$\mu$m emission is spectroscopically unresolved or with FWHM systematically narrower than that of the NLR forbidden lines in Seyfert galaxies. This was interpreted in terms of a kinetic disconnection between the molecular and the NLR gas. Further support to this hypothesis comes from Integral Field Unit (IFU) studies of AGNs, which suggest that the \h2\ is in the galaxy plane, from the centre up to the field limit ($\sim$ 500 pc), while the ionised gas is observed up to high latitudes \citep{rogemar_eso06,rogemar_4051_08,rogemar_4151_09,rogemar_4151_10,rogemar_1066_11,rogemar_1157_11,davies09,mullersanchez09}.

In order to study the \h2\ and \fe2\ lines in LINERs and SFGs, and to complete our analysis of these transitions in emission-line objects, we 
have followed the methodology of our previous studies.  We assume that the widths of the forbidden lines (or NLR gas in the case of AGNs) reflect the large scale motions of the emitting clouds in the gravitational potential of the central mass. As a consequence, similar FWHM of molecular and atomic forbidden lines 
would indicate that these species are co-spatial\footnote{It is worth mentioning that, only because two lines have similar FWHMs does not prove that they are co-spatial. For instance, they could originate approximately in the same region but from different gas layers with different physical conditions. Even when lines originate from exactly the same gas, different line profiles (or FWHM) may result from radiative transfer effects.}.

Table~\ref{width_fwhm} lists the intrinsic FWHM of [S\,{\sc iii}]~0.9531$\mu$m, \fe2 ~
1.2567$\mu$m and H$_{2}$~1.957,~2.0332, and  2.121$\mu$m. To obtain the intrinsic FWHM, the instrumental width was subtracted from the observed line width in quadrature. The
errors in FWHM ($\sim$ 30~\kms) are dominated by the uncertainty in the continuum placement. To ensure that the intrinsic FWHM is equal or larger than the 
instrumental one (360~\kms), a line was considered to be spectroscopically
resolved if its measured FWHM was larger than 500~\kms. 
Thus, lines with FWHM equal to 360~\kms\ in Table~\ref{width_fwhm} may have measured values in the range
360~\kms\ $\leq$ FWHM $\leq$ 500~\kms. 

\begin{figure}
\includegraphics[width=9cm,height=7cm]{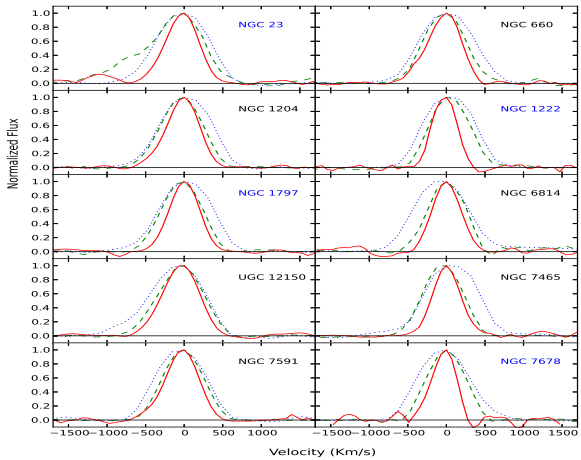} 
\caption{Observed line profiles of [S {\sc iii}] 9531\AA\ (dotted blue line), \fe2 ~1.257$\mu$m 
(dashed green line) and \h2 ~2.122$\mu$m (full line). SFGs are labeled in blue color.\label{prof}}
\end{figure}

Figure~\ref{prof} compares the observed line profiles of [S {\sc iii}] 9531\AA\ (dotted line), \fe2 ~1.257$\mu$m 
(dashed line), and \h2 ~2.122$\mu$m (full line) for a sub-sample of our sources (those exhibiting all the three lines). 
Clearly, [S {\sc iii}] and \fe2\ tend to be broader than \h2, following 
the same trend observed in Seyfert galaxies \citep{ara04,ara05}.

\begin{table}
\centering
\caption{FWHM (in \kms) corrected for instrumental resolution, this correction was applied for lines with FWHM $>$ 500~\kms. Lines with smaller values were reported as unresolved (FWHM=360~\kms).\label{width_fwhm}}
\begin{tabular}{lccccc}
\hline
\hline
Source    &[S {\sc iii}] &[Fe {\sc ii}] &\h2       &\h2       &\h2 \\ 
          &9530          &1.2570        &1.9570    &2.0332    &2.1213 \\ 
          &$\AA$         &\mc           &\mc       &\mc       &\mc \\ 
NGC 23	  & 435 & 563 &  492 &   360 &  360 \\ 
NGC 520	  & --  & --  &  360 &   360 &  360 \\ 
NGC 660	  & 612 & 402 &  390 &   360 &  360 \\ 
NGC 1055  & --  & --  &  --  &   --  &  -- \\ 
NGC 1134  & --  & --  &  --  &   --  &  -- \\ 
NGC 1204  & 562 & 301 &  360 &   360 &  360 \\ 
NGC 1222  & 500 & 447 &  601 &   360 &  360 \\ 
NGC 1266  & --  & 632 &  360 &   360 &  360 \\ 
UGC 2982  & --  & --  &  648 &   360 &  360 \\ 
NGC 1797  & 573 & 285 &  360 &   360 &  360 \\ 
NGC 6814  & 717 & 583 &  368 &   360 &  360 \\ 
NGC 6835  & --  & 499 &  876 &   360 &  360 \\ 
UGC 12150 & 611 & 508 &  527 &   360 &  360 \\ 
NGC 7465  & 557 & 492 &  360 &   390 &  360 \\ 
NGC 7591  & 669 & 367 &  666 &   363 &  360 \\ 
NGC 7678  & 746 & 360 &  360 &   360 &  360 \\ 
\noalign{\smallskip}						       
\hline								       
\noalign{\smallskip}
\end{tabular}
\begin{list}{Table Notes:}
\item The uncertainty associated with the FWHM is $\sim$30 km~s$^{-1}$ (for details see text).
\end{list}

\end{table}

\begin{figure}
\includegraphics[width=8cm]{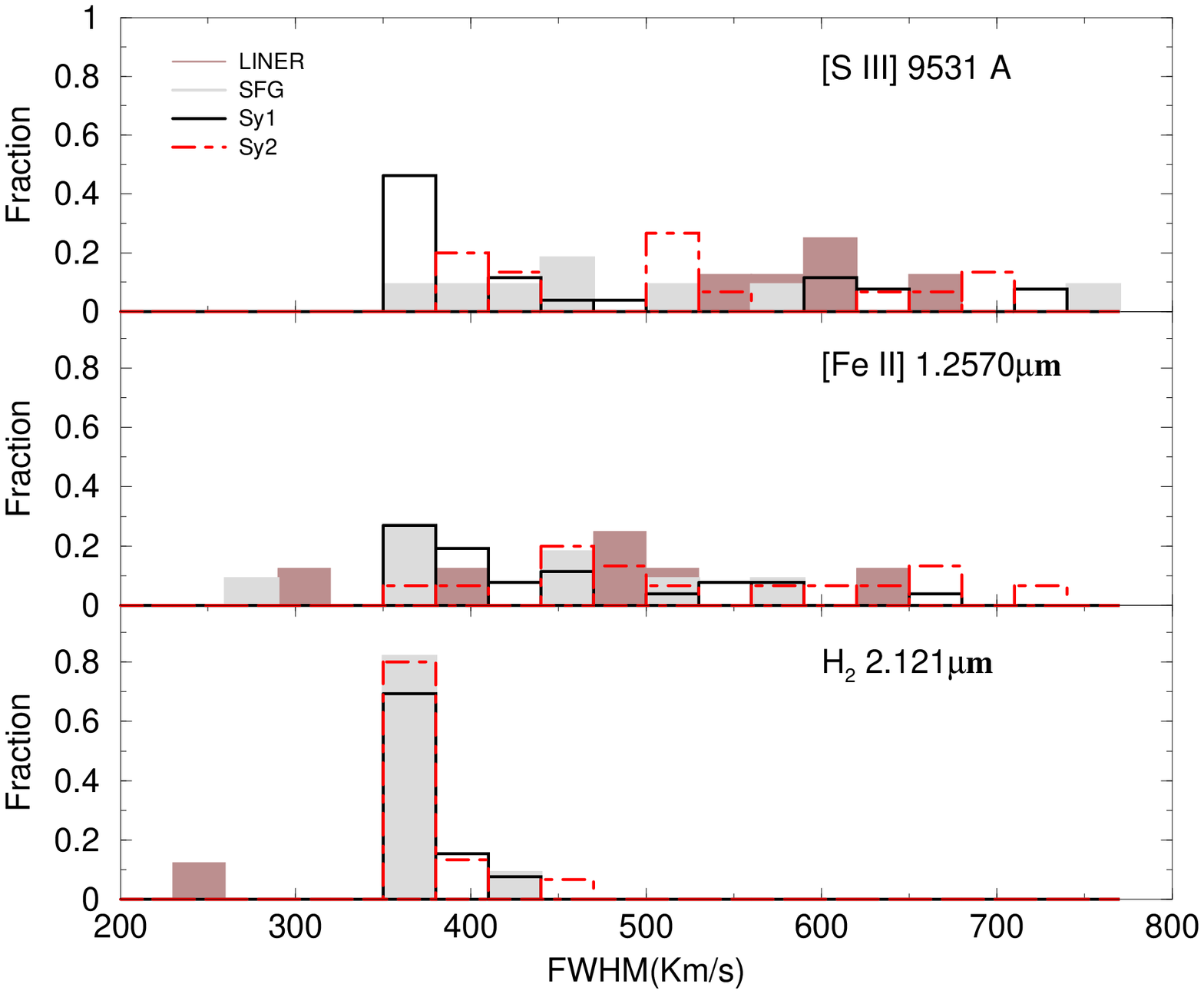} 
\caption{Histogram showing the FWHM distribution for LINERs, SFGs and Sys \citep[the latter from ][]{ara04,ara05}. The measurements for Seyferts were taken from \citet{ara04,ara05}.\label{fwhm}}

\end{figure}

To better understand the kinematics of the gas emitting the different lines in the sources of different classification, 
we show in Figure~\ref{fwhm} a fractional histogram of the FWHM values of the three lines used in Figure~\ref{prof} for the data presented in Table~\ref{width_fwhm} plus literature data taken from \citet{ara04,ara05}.
This plot shows that the molecular gas follows different kinematics than that followed by the atomic forbidden emission-line gas not only in Seyferts, but also in LINERs and SFGs. 
Previous work supports this disconnection in Seyfert galaxies, and suggests that the gas is arranged in a disc surrounding the nuclear region \citep{rka02,rka03}. 
Further support for this hypothesis comes from Gemini-IFU studies \citep{rogemar_eso06,rogemar_4051_08,rogemar_4151_09,rogemar_4151_10,rogemar_1066_11,rogemar_1157_11}, who suggest that the \h2\ is located in the galaxy plane and is distributed along the whole field, while the ionised gas is observed up to
high latitudes from the galaxy plane.

In summary, the observational evidence presented here confirms that \h2\ is
common within the inner few hundred parsecs of emission-line galaxies, regardless of type. The molecular gas follows different kinematics than that of the ionised gas, suggesting that the two emissions are not co-spatial. The \h2\ lines are narrower than the forbidden ones, in particular [S {\sc iii}]. A possible
explanation involves atomic gas outflow. Indeed, \citet{rogemar_12} have shown very recently that the velocity dispersion of gas restricted to the plane (\h2\ gas) does not correlate with that of bulge stars. On the other hand, the velocity dispersion of gas extending to higher latitudes (such as the \fe2\ emitting gas) is similar to that of the galaxy bulge stars. Higher velocity dispersions of \fe2\ relative to stars are probably due to an extra heating provided by a nuclear AGN outflow. In such a scenario, the atomic gas could be farther out from the supermasive black hole (SMBH) than the molecular gas. Another possibility worth mentioning is the atomic gas being closer to the SMBH than the molecular gas and, thus, being more affected by the gravitational pull of the putative SMBH in AGNs. Such a possibility is related to the fact that [S {\sc iii}] is a higher ionisation line and, thus, its bulk should be formed more inwards in the NLR than \fe2\ and \h2. Thus, the unresolved FWHM of \h2\ in almost all objects implies that the molecular gas is probably not gravitationally bound to the SMBH, but to the collective gravitational potential of the galaxy. According to literature results \citep{rka02,rka03}, it may be arranged in a disc-like structure. Further support of this hypothesis comes from the fact that, compared to SFGs, AGNs tend to have significantly broader [S {\sc iii}] than \fe2\ and \h2\ (Fig.~\ref{fwhm}).

\section{Excitation mechanisms of the \h2\ lines}\label{exit}

\begin{figure*}
\includegraphics[width=17.5cm]{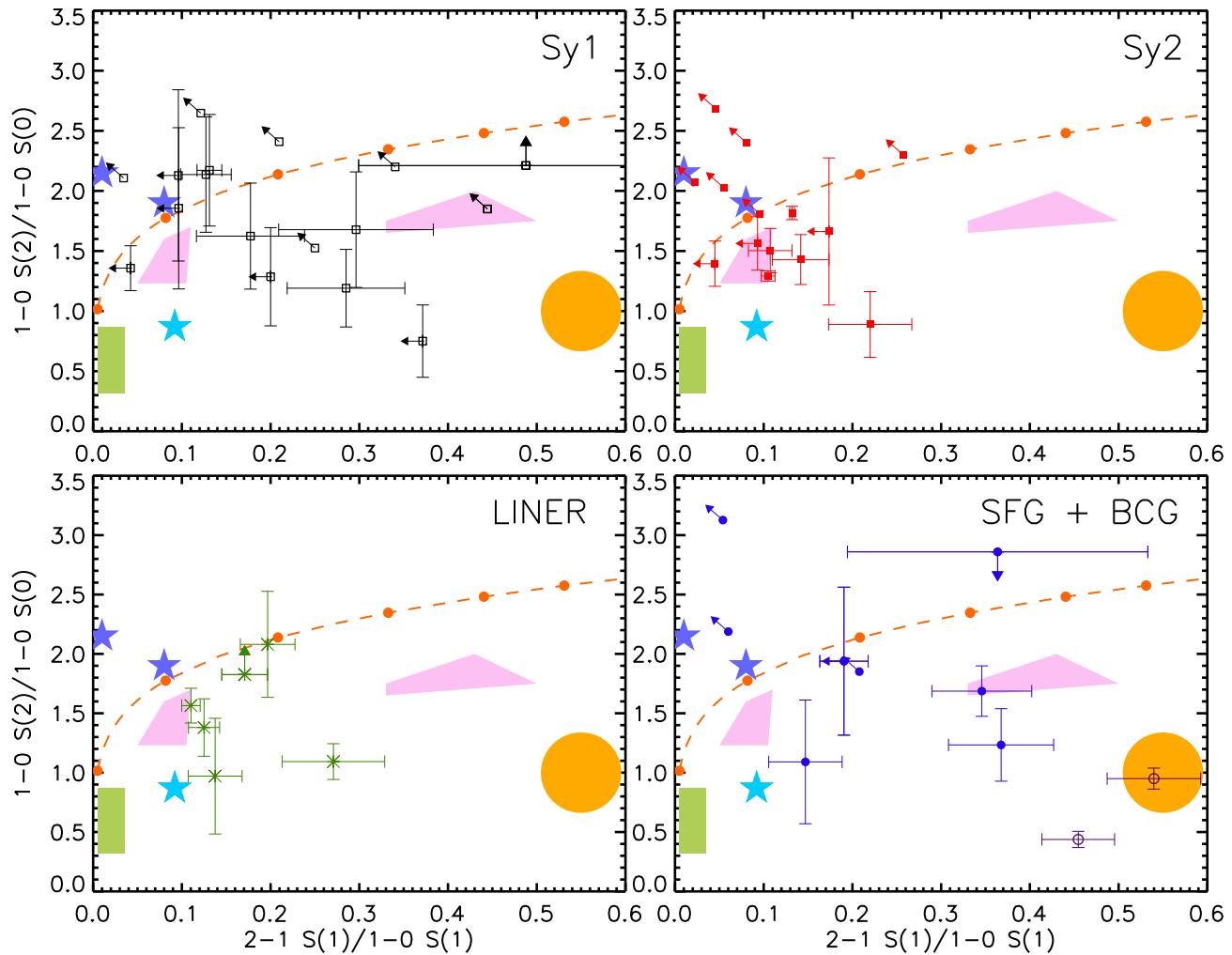}
\caption{Objects with different nuclear activity populate different regions of the \h2\ 1-0~S(2) 2.033$\mu$m/1-0~S(0) 2.223$\mu$m versus 2-1~S(1) 2.247$\mu$m/1-0~S(1) 2.121$\mu$m diagnostic diagram. Different types of nuclear activity are shown in each plot as follows: Sy1, QSO, and NLSy1 (black open squares); Sy2 (red filled squares); LINERs (green asterisks); SFGs (blue filled circles) and BCGs (purple open circles). The orange dashed curve represents the ratios for an isothermal and uniform density gas  distribution; each dot represents the temperatures from 1000 to 6000 K, in steps of 1000 K, from left to right. The orange circle on the right covers the locus of the non-thermal UV excitation models of \citet{bvd}, while the green box (on the left) covers the locus of the thermal UV excitation models of \citet{sd89}. Purple stars are thermal X-ray models of \citet{lm83} and cyan star is a model of shock by  \citet{Kwan_1977}. \citet{oli12} are indicated by the two pink polygons.}

\label{MOURI_OBS}
\end{figure*}

As mentioned in the Introduction, \h2\ line ratios are used to discriminate among the dominant line excitation mechanisms. For example, \citet{mo94} suggested the use of the ratios 1-0~S(2) 2.033$\mu$m/1-0~S(0) 2.223$\mu$m versus 2-1~S(1) 2.247$\mu$m/1-0~S(1) 2.121$\mu$m to separate the mechanism exciting the \h2\ emission. Figure \ref{MOURI_OBS} shows this plot for the objects listed in Table~\ref{log} plus AGNs from \citet{ara04,ara05} and blue compact dwarf galaxies (BCGs) from \citet{izotov11}. It can be seen from Fig. \ref{MOURI_OBS} that objects with different nuclear activity populate different regions of the diagram. Sy1s and SFGs tend to scatter over the plotted region, although Sy1s tend to have lower values of 2-1~S(1)/1-0~S(1) than SFGs, while Sy2 and LINERs are concentrate in a smaller region around the coordinates (0.1,1.5). Note also that two BCGs are on the lower right of the plot.

Figure \ref{MOURI_OBS} also shows results of model calculations found in the literature. The UV-excited low-density gas models of \citet{bvd} occupy the region indicated by the orange filled circle on the right, while the UV-excited high-density models of \citet{sd89} occupy the region indicated by the green box on the left. Models of X-ray irradiated gas of \citet{lm83} are indicated by purple stars. The cyan star is a model of shock by \citet{Kwan_1977}. \citet{mht96} showed that for X-ray--irradiated molecular gas, the 2-1~S(1)/1-0~S(1) ratio is $\lesssim$0.3 ($\sim$0.1 in regions where the line 1-0 S(1) is more intense). \citet{oli12} calculated the ratios with the photoionisation code CLOUDY for AGN models. They found that heating by X-rays produced by active nuclei is a common and very important mechanism of \h2\ excitation. Shock models by \citet{Kwan_1977} give 2-1~S(1)/1-0~S(1) = 0.092 and 1-0~S(2)/1-0~S(0) = 0.87, while \citet{hm89} find ratios in the range 0.3-0.5 for the 2-1~S(1)/1-0~S(1) ratio\footnote{This ratio is not indicated on Fig. \ref{MOURI_OBS} because the authors did not provide the corresponding 1-0~S(2)/1-0~S(0) ratios.}.

The observed ratios are not reproduced by any of pure low-density UV, high-density UV, X-rays or shock models only. Different processes may be more important in different objects. Moreover, a mixture of these processes are likely to be exciting the molecule in both AGNs and SBs. Here we will focus our discussion in the excitation due to a radiation source (i.e., no shock excitation).

The models of \citet{sd89} and \citet{bvd} only consider UV radiation (E $<$ 13.6 eV). On the other hand, the models of \citet{mht96} only consider X-ray radiation (E $\gtrsim$ 100 eV). \citet{lm83} present very complete models and take a wider energy range into account, but they only calculated two models. \citet{oli12} obtained models for a single value of gas density ($n_H =$ 10$^4$ cm$^{-3}$) and their calculations stop when the temperature falls to $T =$ 1000 K, missing an important part of the \h2\ emission, where the X-rays may be of great importance \citep[see the discussion of Fig.~\ref{MOURI_RR} below and ][]{Aleman_Gruenwald_2011}.

Here we present new calculations of the \h2\ IR emission lines produced with a photoionisation numerical code. These calculations are an improvement over the previous mentioned calculations, since they consider the spectrum from the IR up to X-rays (10$^{-5}\,keV \lesssim E \lesssim$ 1\,keV), they run from the inner and more ionized zones to the outer and more neutral zones, where T decreases to 100K. The processing of the radiation in the inner gas shells is done in a self-consistent way and the microphysics of \h2\ is included in detail. The models are described below (Sect. \ref{modH2}). In Sect. \ref{compH2}, we compare model and observed \h2\ line ratios, with the aim of studying the excitation mechanism of these lines.

\subsection{Models} \label{modH2}

We calculated the intensity of \h2\ emission lines of AGNs using the one-dimensional photoionisation code {\sc aangaba} \citep{Gruenwald_Viegas_1992}. The \h2\ microphysics is included in detail in the code. To calculate the intensity of \h2\ IR lines, the density and the population of the \h2\ electronic ground state rovibrational levels that originate the emission must be known. The density of \h2\ is calculated assuming the chemical and ionisation equilibrium between this molecule and the other H-bearing species (H$^0$, H$^+$, H$^-$, H$_2^+$ , and H$_3^+$; see Aleman \& Gruenwald 2004). The code includes over forty different reactions of formation and destruction of these species. The population of the \h2\ rovibrational levels of the three lowest electronic bound states is calculated by assuming statistical equilibrium (i.e., the total population rate of a level is equal to the total depopulation rate). The populations of the electronic ground level takes into account the radiative and collisional excitation and de-excitation mechanisms, as well as the possibility that \h2\ is produced or destroyed in any given level. For upper electronic levels, only radiative electronic transitions between each upper state and the ground state are included, since this must be the dominant mechanism. 

The population of excited \h2\ rovibrational levels of the electronic ground state occurs mainly by electric dipole transitions to upper electronic states, with the absorption of a UV photon, followed by the subsequent decay to the ground state (with the emission of an UV photon), usually in a excited rovibrational level.  This mechanism of \h2\ excitation is called UV pumping. The decay to lower levels, through quadrupole transitions, produces fluorescence. Collisions with the dominant species of the gas may also be an important route of \h2\ excitation. We included collisions of H$_2$ with the main components of the gas, i.e., H, H$^+$, He, H$_2$, and electrons. 

The \h2\ molecule is also included in the gas temperature calculation, which assumes thermal equilibrium (the total input of energy in the gas is balanced by the total loss of energy per unit time and volume). The relevant mechanisms of gain and loss of energy by the gas due to atomic species, dust, and \h2\ are taken into account. The gas heating mechanisms are photoionisation of atoms, atomic ions, and \h2\ by the primary and diffuse radiation, \h2\ photodissociation (direct and two steps), \h2\ formation on grain surfaces, by associative detachment, and by charge exchange with H, \h2\ collisional de-excitation,  and photoelectric effect on dust surfaces. The gas cooling mechanisms are emission of collisionally excited lines; radiative and dielectronic atomic recombination, thermal collisional atomic ionisation, free-free emission, collisional excitation of \h2\, destruction of \h2\ by charge exchange with H$^+$, H$_2$ collisional dissociation, and collision of gas-phase particles with dust grains.More details on the calculation of the H$_2$ level population are found in \citet{Aleman_Gruenwald_2004,Aleman_Gruenwald_2011}.

The emissivity of a line produced by de-excitation from a given
level is the product of the population of that level by the Einstein
coefficient of the transition and by the energy of the emitted photon.
The emissivities of the \h2\ IR lines are calculated for each
point of the nebula. The intensities are calculated by integrating
the emissivity over the nebular volume. The gas is assumed to
be optically thin for the \h2\ lines.

We assume a spherical distribution, uniform density and composition for gas and dust, and gas densities ($n_H$) from 10$^3$ to 10$^5$ cm$^{-3}$, which are typical for AGNs \citep{os89}. Solar abundance is assumed for the elements in all models \citep{os89}, with values (Table~\ref{Abund_Models}) obtained from \citet{Grevesse_Anders_1989}. The chemical composition affects the processing of the radiation inside the cloud and, consequently, the ionisation structure of the cloud, but this does not significantly affect the \h2\ ratios, as shown by \citet{oli12}.

\begin{table}
\caption{Model elemental abundance relative to H}  
\label{Abund_Models}   
\renewcommand{\footnoterule}{}  
\centering  
\begin{tabular}{c c c c}  
\hline\hline 
Element & Abundance        	& Element & Abundance       	\\
\hline  
H   	& 1.0                  & Mg  	& 3.8$\times$10$^{-5}$\\
He  	& 9.8$\times$10$^{-2}$ & Si  	& 3.5$\times$10$^{-5}$\\
C   	& 3.6$\times$10$^{-4}$ & S   	& 1.6$\times$10$^{-5}$\\
N   	& 1.1$\times$10$^{-4}$ & Cl  	& 3.2$\times$10$^{-7}$\\
O   	& 8.5$\times$10$^{-4}$ & Ar  	& 3.6$\times$10$^{-6}$\\
Ne  	& 1.2$\times$10$^{-4}$ & Fe  	& 4.7$\times$10$^{-5}$\\
\hline  
\end{tabular}
\end{table}

Dust, in our models, is assumed to be composed by spherical graphite grains with optical properties taken from \citet{Draine_Lee_1984, Laor_Draine_1993}. Models were calculated for two different grain radii, 10$^{-2}$ and 10$^{-1}$ $\mu$m, typical interstellar values  according to \citep{Li_2007} and \citet{mathis77}. As shown in \citet{Aleman_Gruenwald_2004}, the dust particle size is more important for the H$_2$ density than the grain type. The dust-to-gas mass ratio ($M_d/M_g$) is assumed constant in all models. Models were calculated for $M_d/M_g =$ 10$^{-3}$ and 10$^{-2}$, average values for planetary nebulae and the interstellar medium \citep{Tielens_2005,Lenzuni_etal_1989,Stasinska_Szczerba_1999}.
 
In our models, the spectrum of an AGN is approximated by a power law, i.e., $L = C E^{-\alpha}$, where $E$ is the photon energy, $L$ is the number of photons per unit time per unit of energy emitted by the source, and $\alpha$ is the spectral index. The constant $C$ roughly indicates the radiation field intensity. It is usually defined in terms of the ionisation parameter, which is the ratio of the amount of hydrogen ionising photons incident on the inner border of the cloud to the density of H atoms of the cloud. The ionisation parameter ($U$) is then
 
\[
U = \frac{Q_H}{4\pi R_{in}^2n_Hc}
\]
where $n_H$ is the numerical H density of the cloud, $R_{in}$ is the internal radius of the cloud, $Q_H$ is the number of hydrogen ionising photons emitted by the ionising source per unit of time, and $c$ is the speed of light. $U$ values between 10$^{-4}$ and 10$^{-1}$ are inferred from observations of the NLR of AGNs \citep[e.g.][]{os89}. In this paper, we use models for average $U$ values (0.1, 0.01 and 0.001). The shape of the UV to X-ray SED can differ among AGNs \citep{prieto10}, a range well represented by spectral indices from 1.0 to 1.5, which we use in our models.

For comparison, we also include models with a starburst-like spectrum. For such models, we assume that the ionising spectrum has the shape given by \citet{rogerio_sb_08}, which represents the underlying stellar population of the Starburst (SB) galaxy NGC\,7714. We scale the spectrum to study different number of ionising photons. The models range from ${\rm log(} Q_H{\rm )} =$ 52 to 54, to cover SB values \citep{claus95}. We study the same range of gas and dust parameters as for AGNs.  
 
Our models show that the bulk of the \h2\ near-IR emission in AGNs is produced in the warm region of the gas, so we use T $\gtrsim$ 100 K as the stop criterion for the models. For SBs, there is a non-negligible contribution from gas with T $<$ 100 K in some cases, so the stop criterion is T $\gtrsim$ 50 K.

\subsection{\h2\ Diagnostic Diagram: observations {\it versus} models} \label{compH2}

\begin{figure*}
\includegraphics[width=17.5cm]{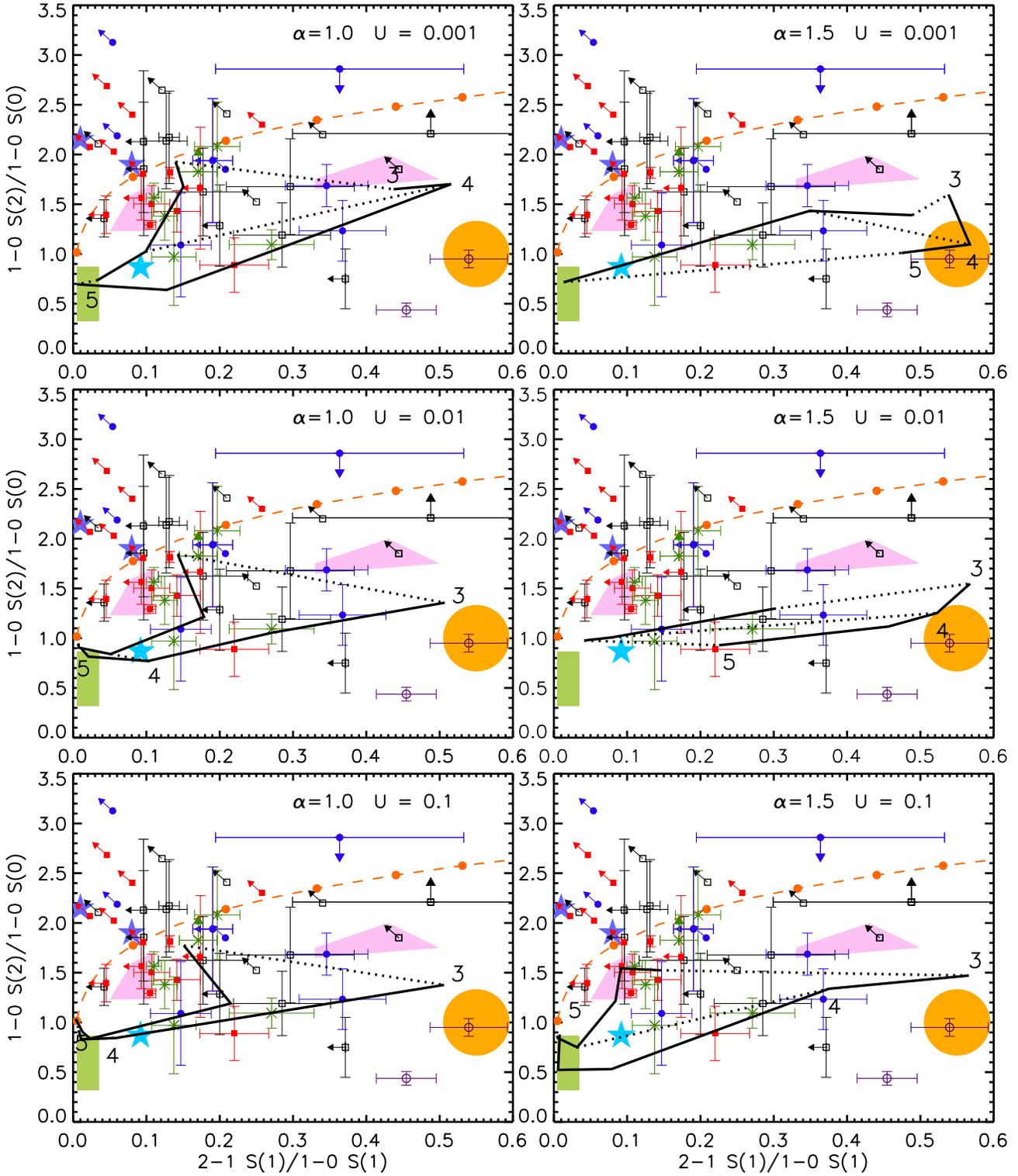}
\caption{\h2\ diagnostic diagram: comparing AGN models and observations. Symbols represent observations, the polygons represent models found in the literature, and the dashed curve represents the ratios for an isothermal and uniform gas density distribution; they follow the same notation as in Fig. \ref{MOURI_OBS}. Calculated ratios are show in each panel as solid lines. Each panel shows AGN models for a different power law spectra, with the ionisation parameter $U$ and the spectral index $\alpha$ as indicated. Each curve connects models for a dust grain size, $a_g =$ 10$^{-1}$ (left curve) and 10$^{-2} \mu$m (right), where the gas density is varied. Dotted lines connect models of the same density for log $n_H =$ 3, 4, and 5. These models assume that the cloud is at 1 pc from the central source.}
\label{MOURI_1PC}
\end{figure*}

\begin{figure*}
\includegraphics[width=175mm]{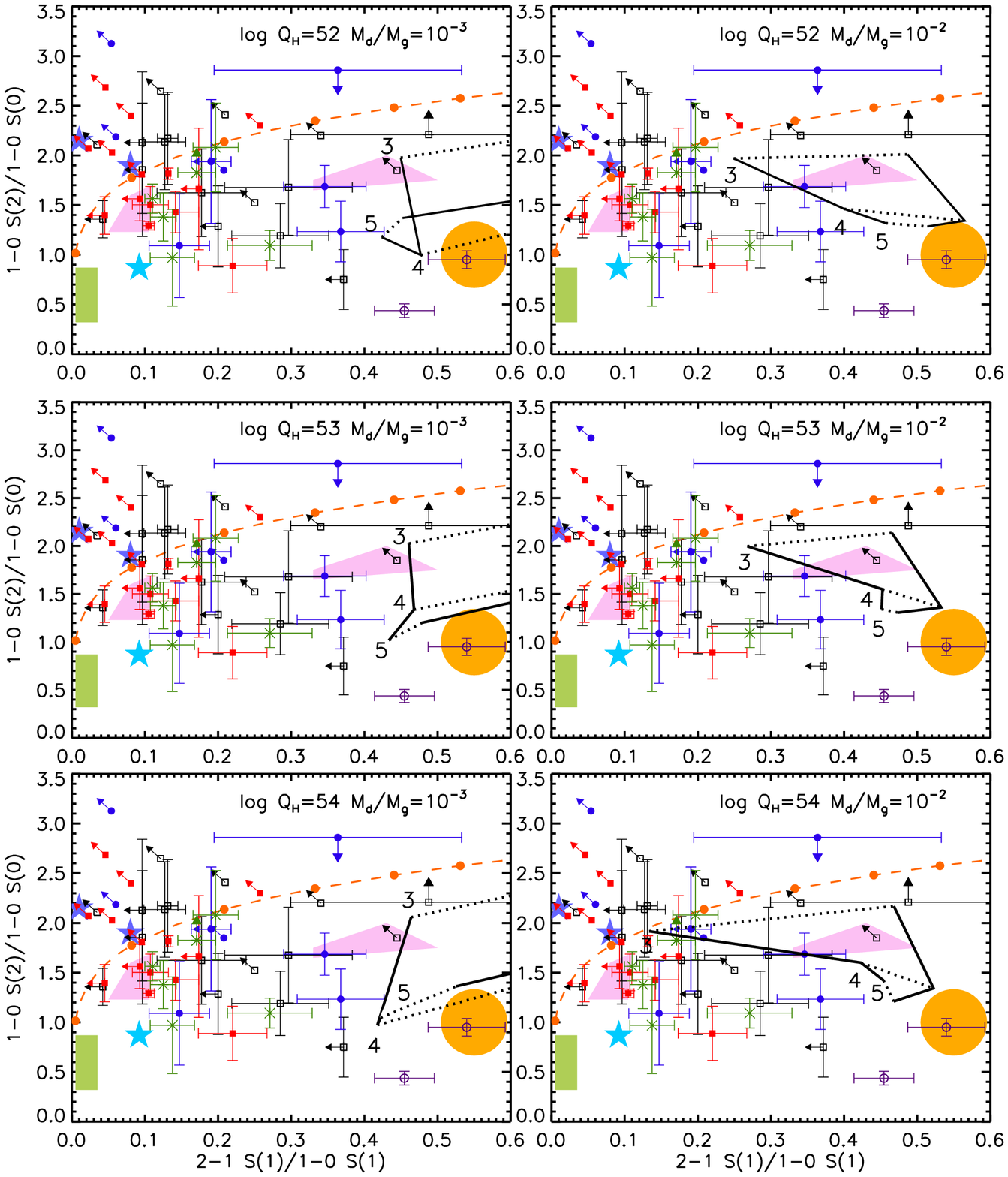}
\caption{Same as Fig. \ref{MOURI_1PC}, but for starburst (SB) models. Calculated ratios are show in each panel as solid lines, connecting models for different gas density. In each panel, the leftmost curve is for models with the dust grain size $a_g =$ 10$^{-1}$ and the rightmost for 10$^{-2} \mu$m. Each panel shows SB models for different number of H ionising photons ($Q_H$) and dust-to-gas ratio ($M_d/M_g$). $Q_H$ is given in photons~s$^{-1}$. Dotted lines connect models of the same density for log $n_H =$ 3, 4, and 5. These models assume that the cloud is at 1~pc from the central source.}
\label{MOURI_SBs}
\end{figure*}

Figure \ref{MOURI_1PC} shows the \h2\ line ratios resulting from our AGN models and from observations. Similarly, Fig. \ref{MOURI_SBs} compares our SBs models with observations. The observational data and the results of previous models showed in Fig. \ref{MOURI_OBS} are also included in the plots of Figs. \ref{MOURI_1PC} and \ref{MOURI_SBs} for comparison.

Our models with typical AGN parameters can reproduce most of the observed values (including the SFGs), while SB models only account for a fraction of the SFGs and BCGs and sub-sample of the AGNs with high 1-0~S(2)/1-0~S(0). Our models confirmed that the \h2\ excitation in AGNs is most likely produced in a gas irradiated by a spectrum rich in high energy photons, while for the SBs the \h2 excitation is could be due to a starburst-like spectra or by a power-law spectra as can be seen in Figs. \ref{MOURI_1PC} and \ref{MOURI_SBs}. 
Models with a hard incident spectrum (i.e., with a large amount of high energy photons, or a lower $\alpha$ value) tend to have lower 2-1~S(1)/1-0~S(1) ratios than AGN models with soft spectra or SB models. The region around the coordinates (0.15,1.5) on these plots is densely populated with Seyfert galaxies and LINERs. For $\alpha = $1.0, only models with $n_H <$ 10$^{4}$ cm$^{-3}$ can reproduce the observations. Models with $\alpha = $1.5 can reproduce the observations, with the exception of the dense ($n_H \sim $ 10$^{5}$ cm$^{-3}$) $U =$ 0.1 models. Alternatively, if the cloud is dense ($n_H \sim $ 10$^{5}$ cm$^{-3}$), the cloud is thiner than our models, i.e., has only gas at temperatures higher than 500K (see paragraph bellow). The observations around the coordinates (0.12,1.5) can only be reproduced by models with $n_H \sim$ 10$^{3}$cm$^{-3}$. It is worth mentioning that the \h2\ line ratios remain essentially unchanged if we consider
an average AGN ionizing spectrum like the one used by \citet{korista97}, or an equivalent power-law. For the latter we mean a SED that produces approximately the same far-UV (5eV) to X-ray (2keV) flux ratio. The \citet{korista97} SED, for example, is equivalent to a single power-law with a spectral index of 1.4. As an example, one of our models with a spectral index of 1.5 gives the ratios 1-0 S(2)/ 1-0 S(0) = 1.252 and 2-1 S(1)/ 1-0 S(1) = 0.523, while a model with the same parameters plus the Korista et al. (1997) SED gives very similar results for 1-0 S(2)/ 1-0 S(0) = 1.174 and 2-1 S(1)/ 1-0 S(1) = 0.566.

\begin{figure}
\includegraphics[width=9cm]{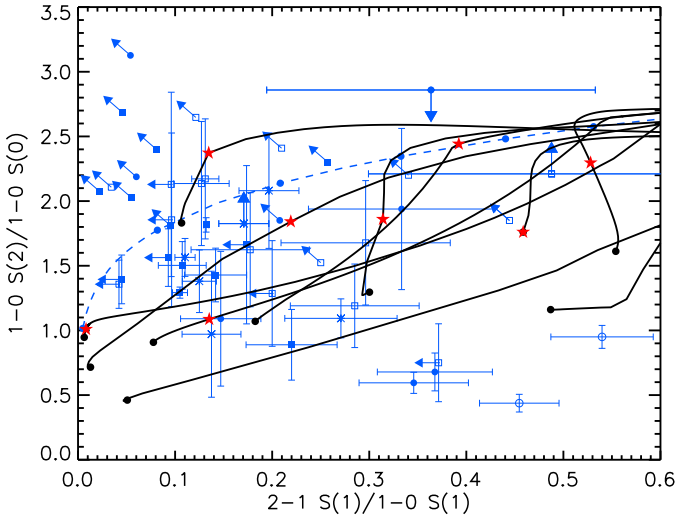}
\caption{Behavior of the ratio 1-0~S(2)/1-0~S(0) vs 2-1~S(1)/1-0~S(1) with the distance from the inner border of the cloud, for several AGN models. Each solid curve represents a model, with distance increasing from right to left. The red stars and black dots over these curves indicate the stop criterion of T = 1000 K and 100 K, respectively. Dots are observations and the dashed curve represents the ratios for an isothermal and uniform gas distribution as shown in Figs. \ref{MOURI_1PC} to \ref{MOURI_SBs}.}
\label{MOURI_RR}
\end{figure}

The increase of the depth in the cloud also affects significantly the line ratios as can be seen in Fig. \ref{MOURI_RR}. In general, for increasing depth both ratios decrease towards lower values, initially following the LTE high temperature solutions (dashed curve) and then following a path that depends on object properties. It can clearly be observed in this figure 
that the ratios are significantly affected if the models run only for $T \geq $ 1000 K (marked with red stars), thus, affecting model predictions.

The diagrams show that several factors, besides the radiation source, compete for the determination of \h2\ line ratios. Gas and dust density, grain size, and the shape and luminosity of the central source spectra are the most important. For example, Figure~\ref{MOURI_1PC} shows that AGN models with high gas density tend to have low 1-0~S(2)/1-0~S(0) and 2-1~S(1)/1-0~S(1) ratios, reflecting an emission spectrum dominated by collisions, as discussed by \citet{sd89}. Values obtained by \citet{Kwan_1977} for shock regions are similar to our densest AGN models. For SBs, high density models also tend to have low values of 1-0~S(2)/1-0~S(0), but the exact behaviour of 2-1~S(1)/1-0~S(1) with density also depends on other properties, as discussed below.

Since we assume a constant dust-to-gas ratio (and not the density of grains), increasing the gas density also increases proportionally the amount of dust. As seen in Fig. \ref{MOURI_SBs}, increasing just the dust-to-gas ratio (keeping $n_H$ constant) diminishes the 1-0~S(2)/1-0~S(0) ratio, but not necessarily 2-1~S(1)/1-0~S(1). We show only SB models with different dust densities, but the effect of increasing dust density on AGN models is similar to that on SB models.  

As previously mentioned, dust grain size also plays an important role in \h2\ line ratios. Models with smaller dust grains have smaller 2-1~S(1)/1-0~S(1) ratios. Assuming a constant dust-to-gas ratio, the optical depth due to dust and the \h2 formation rate are inversely proportional to grain size, as shown in \citet{Aleman_Gruenwald_2004}. For these reasons, more \h2\ is produced in models with smaller grains.

\subsection{\h2\ excitation temperatures and masses}

As stated in \citet{ara05}, an alternative way to determine the mechanism driving the molecular gas excitation is to 
derive the rotational and vibrational temperatures for the \h2. A gas dominated by thermal excitation has 
similar rotational and vibrational temperatures (as expected for a gas in LTE) while fluorescent excitation is characterized by a high vibrational temperature and a low rotational
temperature. For more details see \citet[][and references therein]{ara05}.

The values of $T_\mathrm{vib}$ and $T_\mathrm{rot}$ can be calculated using the fluxes of the observed
\h2\ lines together with the expressions \citep{rka02}: $T_\mathrm{vib} \cong 5600/\mathrm{ln}(1.355\times
I_\mathrm{1-0S(1)}/I_\mathrm{2-1S(1)})$ and  $T_\mathrm{rot} \cong
-1113/\mathrm{ln}(0.323\times I_\mathrm{1-0S(2)}/I_\mathrm{1-0S(0)})$. The derived values are presented in Table~\ref{tempMass}. It can be observed from that table and from \citet{ara05} Table~4, that AGNs and LINERs tend to have similar values of $T_\mathrm{vib}$ and $T_\mathrm{rot}$, while in the case of SFGs (NGC~23, NGC~520, NGC~34, NGC~1614 and NGC~7714, the latter 3 are from Rodrigues-Ardila sample) the $T_\mathrm{vib}$ tends to be higher than $T_\mathrm{rot}$.

In addition to the excitation temperature, the fluxes of Table~\ref{fluxes} also allow us to compute the mass of hot \h2\ emitting the 1-0S(1)~2.121~$\mu$m line by means of the equation: $m_{\rm H_{2}} \simeq 5.0875 \times 10^{13} D^{2}I_{1-0S(1)} $ \citep{rka02}.

The mass of this hot \h2\ for our galaxy sample is listed in Table~\ref{tempMass}.  They were calculated assuming $T$=2000~K, a  transition probability $A_{S(1)}$=3.47$\times$10$^{-7}$~s$^{-1}$
\citep{tur77}, the population fraction in the $\nu$=1, $J$=3 level $f_{\nu=1, J=3}$=0.0122 \citep{sco82} and the intrinsic flux of \h2\ 2.121\mc, $I_{1-0S(1)}$. The extinction coefficient, C$_{ext}$, is calculated by assuming an intrinsic value of 5.88 for the flux ratio Pa$\beta$/Br$\gamma$ \citep[][, case B]{os89}.
As can be observed in Table~\ref{tempMass} and in \citet{ara05}, the mass of the hot \h2\ is very similar for all activity types.  The fraction of molecular mass present in the nuclear region and emitting in the NIR is a very small fraction of the warm molecular mass expected to be present in the galaxy centre \citep[up to 10$^{10}M\odot$, ][ for example]{young91}, and is not related to activity type.

Overall, the diagnostic diagrams presented in Figs. \ref{MOURI_1PC} and \ref{MOURI_SBs} allow us to distinguish fairly well between the \h2 excitation mechanisms (collisional or UV pumping). However, it is not straightforward to distinguish between different classes of sources (e.g. AGN or Starburst), as discussed in the previous subsection. Alternatively mechanisms may dominate the \h2 excitation in different regions in a same object. As the emission lines are integrated along the line of sight, we observe a combination of such processes. As can be seen in Figs. \ref{MOURI_1PC} and \ref{MOURI_SBs}, both collisional and UV pumping excitations may play an important role not only in AGNs but also in SFG galaxies. This is further supported by the similarity between the vibrational and rotational temperatures of \h2\ in some objects, and the tendency of $T_{\rm vib}$ to be higher than $T_{\rm rot}$ in others (see Table~\ref{tempMass}). For example, in SFGs, the thermal processes may dominate the excitation of the gas irradiated by supernova remnants (SNRs) or very hot stars. In this case, the \h2\ emission is due to X-ray excitation, which would favour vibrational transitions over rotational transitions \citep{ara05}. \citet{mo94} showed that SNRs lies very close to the thermal curve in these diagnostic diagrams (see their Fig.~1). In the case of UV excitation by stars, the effective optical depth of the nebula, a non-thermal signature, will be determined if the ratios reflect the domination of thermal or non-thermal process (Sternberg \& Dalgarno 1989).

\begin{table*}
\centering
\caption{Fluxes of atomic and molecular lines for our sample, in units of $\rm 10^{-15}\ ergs\ cm^{-2}\ s^{-1}$, measured in the Sample. \label{fluxes}}
\begin{tabular}{lccccccccccc}
\hline
\hline
Source    & [Fe{\sc ii}]   &    Pa$\beta$      &    [Fe{\sc ii}]   &      H$\rm _2$    &     H$\rm _2$   &   H$\rm _2$      &        H$\rm _2$ &     H$\rm _2$    & Br$\gamma$         \\
          &   1.2570\mc    &    1.2820\mc      &      1.6444\mc    &      1.9570\mc    &     2.0332\mc   &     2.1213\mc    &       2.2230\mc  &     2.2470\mc    &  2.1650\mc          \\
NGC 23	  &   10.50\pp0.87  &	     --        &    14.20\pp3.46    &   20.20\pp4.60    &  4.94\pp0.57	 &    10.00\pp1.26   &	4.53\pp2.10 (t)&    1.47\pp0.37   &	0.00\pp0.00	\\ 
NGC 520	  &       --       &	     --        &    3.87\pp0.14    &   2.41\pp0.39     &  1.13\pp0.10	 &    2.40\pp0.18   &	 0.67\pp0.06   &    0.83\pp0.12   &	6.67\pp0.19	\\ 
NGC 660	  &   13.90\pp0.65  &	 29.0\pp0.60   &    15.70\pp0.90    &   8.87\pp0.80     &  3.67\pp0.56	 &    6.91\pp0.77   &	 $<$2.01       &  1.18\pp0.12 (t) &	16.60\pp0.71	\\ 
NGC 1055  &       --       &	     --        &        --         &       --          &  0.00\pp0.00	 &        --        &	     --        &        --        &	--       	\\ 
NGC 1134  &       --       &	     --        &        --         &       --          &  0.00\pp0.00	 &        --        &	     --        &        --        &	--      	\\ 
NGC 1204  &   5.16\pp0.2   &	 12.10\pp0.20   &    6.51\pp0.36    &   4.17\pp0.51     &  1.81\pp0.34	 &    3.61\pp0.25   &	 0.87\pp0.09   &    0.71\pp0.10   &	5.88\pp0.27	\\ 
NGC 1222  &   6.32\pp0.43  &	 28.20\pp0.37   &    4.65\pp0.18    &   1.95\pp0.34     &  0.95\pp0.14	 &    0.84\pp0.12   &	 0.49\pp0.14   &    $<$0.16       &	6.98\pp0.073	\\ 
NGC 1266  &   3.19\pp0.51  &	 0.75\pp0.08   &    2.90\pp0.38    &   15.10\pp0.47     &  5.10 \pp0.44	 &    13.70\pp0.25   &	 3.26\pp0.12   &    1.51\pp0.14   &	1.40\pp0.33	\\ 
UGC 2982  &       --       &	     --        &        --         &   1.79\pp0.48     & 0.61\pp0.08 (t) &  0.60\pp0.02 (t) &	     --        &        --        &	0.80\pp0.036	\\ 
NGC 1797  &   5.04\pp0.25  &	 12.10\pp0.26   &    4.65\pp0.47    &   4.10\pp0.51     &  1.48\pp0.30	 &    3.21\pp0.28   &	 1.20\pp0.17   &    1.18\pp0.16   &	5.33\pp0.09	\\ 
NGC 6814  &   4.64\pp0.52  &	 3.53\pp0.57   &    5.53\pp0.48    &   3.36\pp0.54     &  1.25\pp0.20	 &    2.14\pp0.20   &	 1.05\pp0.23   &    0.61\pp0.13   &	0.52\pp0.32	\\ 
NGC 6835  &   2.06\pp0.23  &	 4.86\pp0.19   &    3.24\pp0.08    &   3.33\pp0.44     &  0.85\pp0.05	 &    1.34\pp0.21   & 0.65\pp0.18 (t)  &        --        &	4.66\pp0.25	\\ 
UGC 12150 &   4.73\pp0.33  &	 7.86\pp0.30   &    3.56\pp0.18    &   5.25\pp0.05     &  1.65\pp0.07	 &    4.00\pp0.13    & 1.70\pp0.85 (t)  &  0.55\pp0.12 (t) &	2.88\pp0.13	\\ 
NGC 7465  &   11.8\pp0.66  &	 9.57\pp2.61   &    9.20\pp0.28    &   4.33\pp0.16     &  2.36\pp0.39	 &    3.92\pp0.27   &	 1.71\pp0.10   &    0.49\pp0.06   &	3.75\pp0.69	\\ 
NGC 7591  &   6.80\pp0.21  &	 9.74\pp0.21   &    6.05\pp0.60    &   8.17\pp0.26     &  2.46\pp0.20	 &    4.80\pp0.36   &	 2.25\pp0.25   &    1.30\pp0.26   &	4.07\pp0.052	\\ 
NGC 7678  &   3.28\pp0.49  &	 9.20\pp0.51   &    3.09\pp0.21    &   1.45\pp0.10     &  2.03\pp0.20	 &    0.88\pp0.14   &	 $>$0.71       &    0.32\pp0.14   &	2.56\pp0.15	\\ 
\noalign{\smallskip}
\hline
\noalign{\smallskip}
\multicolumn{6}{l}{(t) Affected by telluric absorption.}\\
\end{tabular}

\end{table*}

\begin{table}
\centering
\caption{Extinction coefficient. Molecular gas mass, vibrational and rotational temperatures.\label{tempMass} }
\begin{tabular}{lccccc}
\hline
\hline
Source    & C$_{ext}$  &  H$_{\rm 2}$ Mass & T$\rm _{vib}$ & T$\rm _{rot}$  \\ 
               &                   &  M$\odot$ &  K                   & K    \\
NGC 23	    &  --  &   --	 &    2521\pp469 &  1067\pp601  \\
NGC 520	    &  --  &   --	 &    4115\pp279 &  1832\pp159	\\
NGC 660	    & 0.69 &  161\pp5	 &    2704\pp240 &  2108\pp208  \\
NGC 1055    &  --  &   --	 &	 --	 &	--	\\		       
NGC 1134    &  --  &   --	 &	 --	 &	--	\\		       
NGC 1204    & 0.59 &  1991\pp57  &     2912\pp256 &  2776\pp270  \\
NGC 1222    & 0.21 &  67\pp7	 &     $<$ 2856  &  2373\pp399  \\
NGC 1266    & 1.35 &  7189\pp132 &     2232\pp154 &  1631\pp118  \\
UGC 2982    &  --  &   --	 &	 --	 &	--	\\
NGC 1797    & 0.54 &  1606\pp56  &     4293\pp271 &  1209\pp307   \\
NGC 6814    & 0.00 &  56\pp19	 &     3607\pp381 &  1165\pp343  \\
NGC 6835    & 0.98 &  201\pp6	 &	 --	 &	--	\\
UGC 12150   & 0.43 &  3227\pp97  &     2444\pp363 &  960\pp630   \\
NGC 7465    & 0.47 &  323\pp34   &     2362\pp227 &  1378\pp222  \\
NGC 7591    & 0.51 &  2638\pp82  &     3478\pp361 &  1069\pp174  \\
NGC 7678    & 0.28 &  170\pp17   &     4225\pp806 &  -- 	 \\
\noalign{\smallskip}						       
\hline								       
\noalign{\smallskip}
\end{tabular}
\end{table}

\section{Nuclear activity from Near-Infrared emission-lines}\label{diagnostic_diagrams}

Since \h2\ and \fe2\ lines are common features in a wide variety of sources, it is interesting to investigate how these emission-lines 
compare in objects with different degrees of nuclear activity.
A diagram involving  the line ratios \h2\ 2.121$\mu$m/Br$\gamma$ and
\fe2\ 1.257$\mu$m/Pa$\beta$ was proposed by \citet{lar98} in a study of
LINERs and other emission-line objects, including a few Seyferts. They report a
strong linear correlation in the log-log plot of \fe2/Pa$\beta$ {\it versus} \h2/Br$\gamma$,
with SB galaxies displaying the lower values, Seyferts with intermediate ones and LINERs with the highest ratios.  They suggest that \fe2/Pa$\beta \sim$1 and \h2/Br$\gamma \sim$3 
mark the end of Seyfert-like nuclei and the beginning of LINER-like
objects.  \citet{ara04,ara05} confirm such diagram as a suitable means of separating
emission-line objects by their degree of activity in the NIR, and propose that 
AGNs are characterized by \h2\ 2.121$\mu$m/Br$\gamma$ and \fe2\ 1.257$\mu$m/Pa$\beta$ flux ratios between 0.6 and 2.
Starburst/H\,{\sc ii} galaxies display line ratios $<$0.6, while LINERs are characterized by values higher than 2
for both ratios. However, the lack of an adequate number of objects preclude definitive conclusions in those studies. 

\begin{figure*}
\includegraphics[width=15cm]{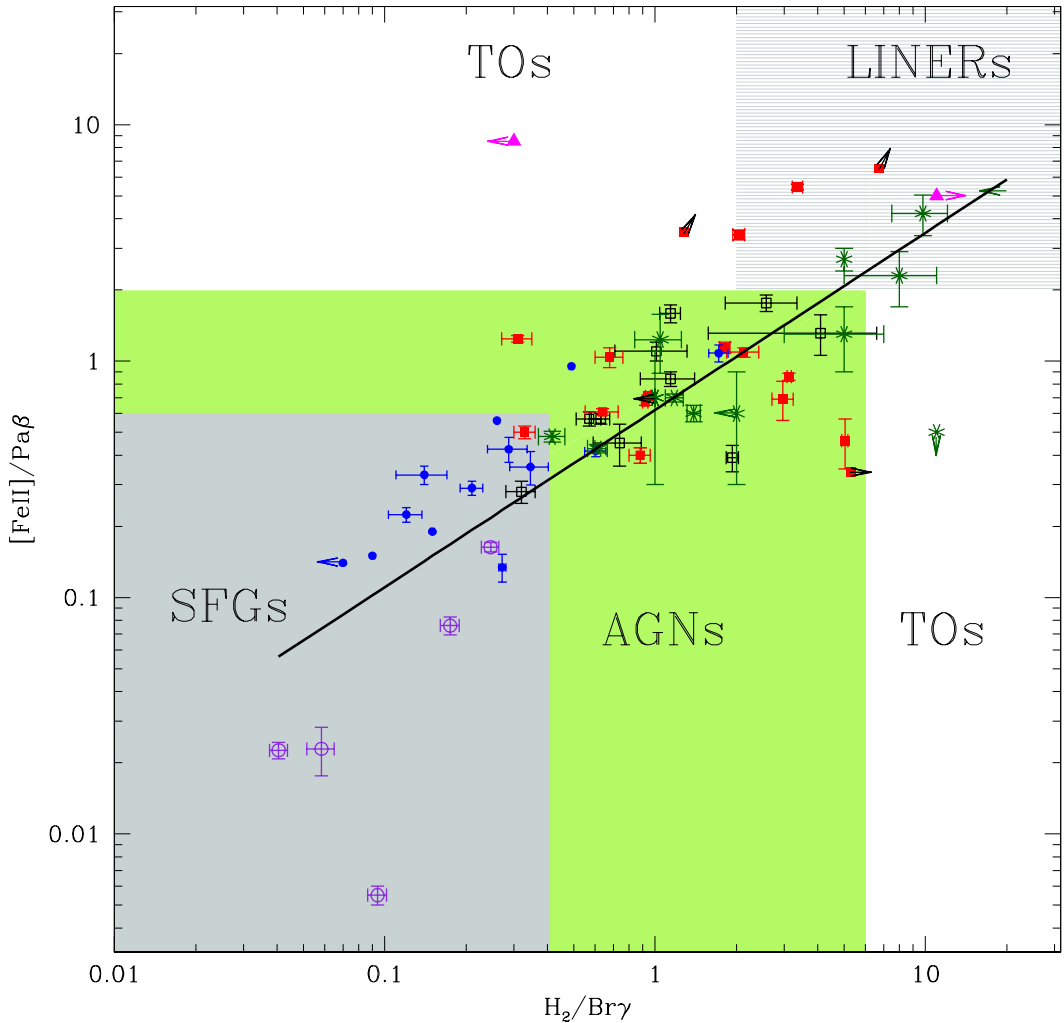} 
\caption{Diagnostic diagram: boxes are AGNs, Sy~1 in black (open) and Sy~2 in red (filled), asterisks are LINERs (green), filed circles are SFG (blue), triangles (magenta) are SNRs and open circles are blue compact dwarf galaxies (violet). The solid line is \fe2/Pa$\beta$ = 0.749($\pm$0.072)\h2/Br$\gamma \rm-0.207\pm0.046$ with correlation coefficient = 0.80 and $r^2$=0.64.}
\label{diag}
\end{figure*}

Here, with a larger sample, and most importantly, a more adequate number of objects of each activity class, we may provide further support to the trend already observed. Figure~\ref{diag} presents the
updated version of \h2\ 2.121$\mu$m/Br$\gamma$ $\times$
\fe2\ 1.257$\mu$m/Pa$\beta$  diagnostic diagram. 
The sample analyzed here is composed by the objects listed in Table~\ref{log} plus some taken from the literature: AGNs from \citet{ara04,ara05}, SFG from \citet{dal04} and \citet{lar98}, LINERs, supernova remnants (SNRs) from \citet{lar98}, and BCGs from \citet{izotov11}. As can be seen in Figure~\ref{diag}, LINERs and AGNs are not clearly separated by such a diagram. However, our sample of LINERs is composed by Luminous Infrared Galaxies (LIRGs), and the classification in Table~\ref{log} is based on optical emission-line diagrams, which are affected by reddening \citep[see ][ for example]{ara05}. Since LIRGs harbour large quantities of dust, we suggest that the LINERs lying within the region of AGNs may hide a Seyfert nucleus.

With our sample comprising $\sim$ 65 objects, we not only
confirm the diagram (Fig.~\ref{diag}) as a discriminator of emission-line objects by their degree of activity, but we also provide improved limits to the line ratios for each activity type. The limits are as follows: 
(i) SFGs \fe2\ 1.257$\mu$m/Pa$\beta \lesssim$ 0.6 and \h2\ 2.121$\mu$m/Br$\gamma \lesssim$ 0.4, 
(ii) AGNs 0.6 $\lesssim$ \fe2\ 1.257$\mu$m/Pa$\beta \lesssim$ 2 and 0.4 $\lesssim$ \h2\ 2.121$\mu$m/Br$\gamma \lesssim$ 6, and
(iii) LINERs \fe2\ 1.257$\mu$m/Pa$\beta \gtrsim$ 2 and \h2\ 2.121$\mu$m/Br$\gamma \gtrsim$ 6,

As discussed in \citet{ara05} on a pure observational basis, the interpretation of Figure~\ref{diag} is not straightforward. For example, the correlation between \h2/Br$\gamma$ and \fe2/Pa$\beta$ may originate from the fact that \h2\ and \fe2\ are mainly excited by the same mechanism, i.e., X-ray. In fact, it was shown very recently by \citet{oli12}, using photodissociation models, that X-ray heating from the AGN is an important mechanism in the excitation of these NIR emission-lines in AGNs (see Sec.\ref{exit}).  They also show that a decrement in the X-ray content of the continuum source translates into a line weakening, and their models are no longer compatible with observations. However, X-ray heating is a plausible explanation for AGNs, but may not work for LINERs.  \citet{lar98}, for example, argue that hard X-ray heating from a power-law is 
a plausible mechanism to explain LINERs with low values of the \fe2/Pa$\beta$ and \h2\ 2.121$\mu$m/Br$\gamma$ ratios. But, it would not explain the high values (up to $\sim$7) observed in some objects. Nevertheless, the high \fe2/Pa$\beta$ line ratios may be explained by the dependence of \fe2/Pa$\beta$ on the Fe/O abundance and on the dependence on the ionisation parameter of the molecular gas, leading to high values for both ratios \citep[][see their Figure~1]{oli12}

\begin{figure*}
\includegraphics[width=15cm]{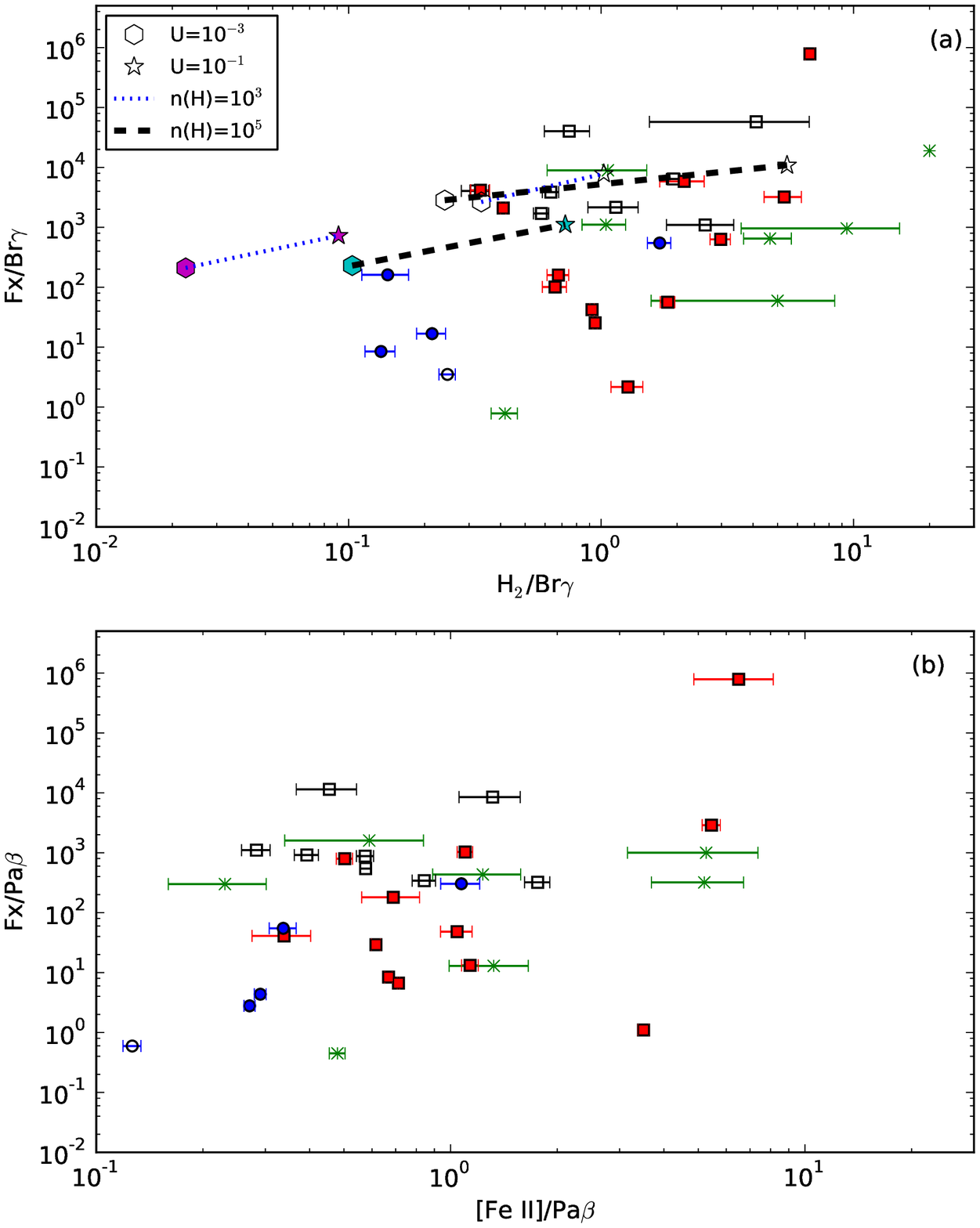} 
\caption{Hard X-ray flux versus emission line ratios. The simblos for the observations are the same as in Fig.~\ref{diag}. Hexagons represent models with U=10$\rm ^{-3}$ and stars with U=10$\rm ^{-1}$ (open symbols are for $\alpha$=1 and filled for $\alpha$=1.5). Points joined by dotted and dashed lines represent n(H)=10$\rm ^3$ and n(H)=10$\rm ^5$, respectively.}
\label{fluxflux}
\end{figure*}

On the other hand, it has been recently proposed that it is unlikely that LINER emission
lines are powered by photoionisation by the low-luminosity AGN they harbour \citep{eracleous10,cid10,yan12}. This would rule out X-ray
heating as the main mechanism driving the excitation of the NIR emission-lines in LINERs.
We investigated the role played by the hard X-ray (2-10 KeV, F$_{x}$) radiation in \fe2\ and
\h2\ emission lines with diagrams involving F$_{x}$/Br$\gamma$ vs \h2/Br$\gamma$ ratios and F$_{x}$/Pa$\beta$ vs \fe2/Pa$\beta$
(Fig.~\ref{fluxflux}). The X-ray fluxes were taken from literature and are listed
in Table~\ref{Fx}. A mild correlation is observed for F$_{x}$/Br$\gamma$ vs \h2/Br$\gamma$ (Fig.~\ref{fluxflux}a), indicating
that the X-ray heating is an important mechanism in the molecular gas excitation. For a
more objective assessment of the X-ray heating in our data, we included model predictions
to Fig.~\ref{fluxflux}a. It is clear that the explored parameter space overlaps the observed line
ratios. In addition, the AGNs are better explained by $\alpha$=1 and $U$=0.1. It is also interesting
that Sy~1s require higher hydrogen gas densities than the other activity types, indicating that the molecular gas in Sy~1s is denser. No correlation is observed
for F$_{x}$/Pa$\beta$, and, \fe2/Pa$\beta$ and as for the diagram invloving \h2/Br$\gamma$ no separation is observed between LINERs and Seyferts
(Fig. ~\ref{fluxflux}). This result suggests that X-ray heating alone is not responsible for the
observed sequence in Fig.~\ref{diag}. As shown by \citet{oli12}, high values of \fe2/Pa$\beta$
line ratios can be explained by the dependence on the Fe/O abundance. Thus, the observed
sequence in Fig.~\ref{diag} may be a consequence of the amount of X-ray photons together with abundance
effect (i.e. in the case of LINERs powered by star-formation, SN explosions would eject
enriched material to the interstellar medium).

A detailed interpretation of the variation of these line ratios, from pure-starburst driven
emission to LINER nuclei, is non-trivial. A possible scenario is that the different
line ratios reflect different excitation mechanisms, in the sense that they are able to
describe a transition from an ionizing radiation powered by star formation to pure shock
excitation driven by SNRs \citep{ximena12}. In such a case, the emission-line ratios
observed in our LINER sample would be a combination of processes, from X-ray heating to
star-formation/evolution. Since most LINERS have strong star formation (Tab.~\ref{log}), the stars
could provide the required electromagnetic or mechanical power to produce the observed line
strengths. However, to clearly investigate this issue, a 3D simulation of the emitting gas
would be needed (i.e. AGN in the center + star-formation in a volume). It is also worth
mentioning that, in the case of star formation, the underlying stellar population would
significantly affect the atomic hydrogen emission-lines, thus affecting the observed
line ratios  \citep{lar98,rogerio_sb_08}.


\begin{table}
\centering
\caption{Hard X-Ray Fluxes, in units of $\rm 10^{-13}\ ergs\ cm^{-2}\ s^{-1}$, taken from literature. \label{Fx} }
\begin{tabular}{lll}
\hline
\hline
Source       &   Fx(2-10 KeV) & Reference  \\
\noalign{\smallskip}						       
\hline								       
\multicolumn{3}{c}{Sy~1} \\
\hline
\hline
Mrk~334        &  80         &  \citet{lu04} \\
NGC~7469       &  290        &  \citet{lu04} \\
NGC~3227       &  75         &  \citet{lu04} \\
NGC~4151       &  900        &  \citet{lu04} \\
Mrk~766	       &  300        &  \citet{lu04} \\
NGC~4748       &  30         &  \citet{landi10} \\
NGC~5548       &  430        &  \citet{lu04} \\
NGC~6814       &  300        &  \citet{turner89} \\
\noalign{\smallskip}						       
\hline								       
\multicolumn{3}{c}{Sy~2} \\	     
\hline				     
ESO428-G014   & 3.8  	 & \citet{lu04} \\
NGC~591       & 2    	 & \citet{guainazzi05} \\
Mrk~573       & 2.8  	 & \citet{guainazzi05} \\
Mrk~1066      & 3.6  	 & \citet{guainazzi05} \\
NGC~2110      & 430  	 & \citet{nandra07}  \\
NGC~7674      & 5    	 & \citet{lu04} \\
NGC~5929      & 79   	 & \citet{lu04} \\
Mrk~1210      & 250  	 &\citet{lu04} \\
NGC~5728      & 13.3 	 &\citet{shu07} \\
Mrk~993	      & 6.1  	 & \citet{guainazzi05} \\
NGC~5953      & 0.06 	 & \citet{guainazzi05} \\
NGC1144       & 1100 	 &\citet{lu04} \\
\noalign{\smallskip}						       
\hline								       
\multicolumn{3}{c}{LINER} \\	     
\hline				     
NGC~5194     & 4.8  	  &	\citet{cappi06}      \\
NGC~7743     & 0.44 	  &	 \citet{gonzales09}      \\
NGC~660      & 0.13 	  &	\citet{filho04} \\
NGC~7465     & 41.5 	  & \citet{guainazzi05} \\
NGC~3998     & 84.2 	  &  \citet{ueda01}      \\
NGC~7479     & 3.9  	  & \citet{ueda05}       \\
NGC~4736     & 19   	  & \citet{ueda01}    \\
\noalign{\smallskip}						       
\hline								       
\multicolumn{3}{c}{SFG} \\	            		 
\hline				            		 
NGC~34       &  39   		 &\citet{lu04} \\
NGC~7714     &  2.6  		 & \citet{ueda05}      \\
NGC~1614     &  3.6  		 & \citet{ueda01}      \\
NGC~3310     &  17.4 		 & \citet{ueda01}      \\
\noalign{\smallskip}						       
\hline								       
\multicolumn{3}{c}{BCDg} \\	     
\hline	     			     
Mrk~930      &	  0.161              &   \citet{rosagonzales09} \\
\noalign{\smallskip}						       
\hline								       
\noalign{\smallskip}
\end{tabular}
\end{table}

Another interesting point is the fact that Br$\gamma$ is produced mostly inside the hydrogen ionised region, while the \h2\ emission originates in a warm semi-ionised region \citep{Aleman_Gruenwald_2011} around the hydrogen recombination and in the photodissociation region (PDR), where H is mostly neutral. 
SFGs typically have lower 1-0 S(1)/Br$\gamma$ ratios than AGNs. This may indicate that the region that produces the \h2\ emission in AGNs is larger than in SBs, if we consider that AGN spectra are harder than SFG's. Harder ionisation spectra may produce a more extended semi-ionised region (Aleman \& Gruenwald 2004, 2007), which favours the formation and emission of \h2. In our models, the ratio \h2\ 1-0 S(1)/Br$\gamma$ is within the range 10$^{-3}$ $<$\h2\ 1-0 S(1)/Br$\gamma$ $<$ 10 (see Sec.~\ref{exit}), with the higher ratios ($>$ 1) occurring in models with $\alpha = $1.0 and the smallest values ($<$ 10$^{-2}$) occurring for $\alpha = $1.5 (in fact, the emission of these last objects are below the detection limit of our observations). Furthermore, denser and/or dustier models typically have higher ratios. A more detailed discussion on these models will be made in a forthcoming work (Aleman et al., in preparation).

\section{Concluding Remarks}\label{fin}

We present NIR spectroscopy (0.8\mc\ --- 2.4\mc) of seven SFGs and nine LINERs to discuss the distribution and excitation of the emitting \h2\ and \fe2\ gas in emission-line galaxies. These data were analysed together with similar AGN data of two previous publications \citep{ara04,ara05}.  The whole data set constitutes the most complete and homogeneous sample of such kind of objects observed in the NIR to date. Our main conclusions are:

\begin{itemize}

\item \h2\ is common within the inner few hundred parsecs of emission-line galaxies, regardless of type. However,
the molecular gas follows different kinematics than that of the ionised gas, suggesting that 
the two emissions are not co-spatial. For example, it is clear that the \h2\ lines are narrower than the forbidden lines, especially for [S {\sc iii}]. This could be interpreted as if the latter is broadened by the gravitational influence of the SMBH in the AGNs. Furthermore, LINERs tend to have [S {\sc iii}] significantly broader than \fe2\ and \h2\ compared to SFGs.
The fact that \h2\ FWHM is unresolved in almost all objects implies that the molecular gas is probably not gravitationally bound to the SMBH but to the gravitational potential of
the galaxy following a different kinematics than that of the classical NLR gas. In addition, IFU data and rotation curves derived for \h2\ and published by other authors for objects in common with our sample support this hypothesis and point to a scenario where \h2\ may be arranged in a disc-like structure on the galaxy plane.

\item We have computed new photoionisation models  for clouds ionised by AGN and stars, which provided fluxes of the \h2 lines. We use a diagnostic diagram with \h2\ IR line ratios to study the excitation mechanisms of \h2.
The thermal excitation plays an important role not only in AGNs, but also in SFG galaxies.  This hypothesis is further supported by the similarity between the vibrational and rotational temperatures of \h2\ in some objects, and the tendency of $T_{\rm vib}$ to be higher than $T_{\rm rot}$ in others. In SFGs, the importance of the thermal excitation may be associated with the presence of SNRs close to the region emitting \h2\ lines.

\item With our extended sample, of 65 objects, we have confirmed that the diagram involving the line ratios \h2\ 2.121\mc/Br$\gamma$ and \fe2\ 1.257\mc/Pa$\beta$ is an efficient tool for separating emission-line objects according to their dominant degree of activity and should be useful for classifying objects with hidden AGNs or highly reddened objects. We also provide improved limits to the line ratio intervals for each activity type, as follows: 
(i) SFGs \fe2\ 1.257$\mu$m/Pa$\beta \lesssim$ 0.6 and \h2\ 2.121$\mu$m/Br$\gamma \lesssim$ 0.4, 
(ii) AGNs 0.6 $\lesssim$ \fe2\ 1.257$\mu$m/Pa$\beta \lesssim$ 2 and 0.4 $\lesssim$ \h2\ 2.121$\mu$m/Br$\gamma \lesssim$ 6, and
(iii) LINERs \fe2\ 1.257$\mu$m/Pa$\beta \gtrsim$ 2 and \h2\ 2.121$\mu$m/Br$\gamma \gtrsim$ 6, 

\item A positive correlation is found in the diagnostic diagram involving the above ratios, interpreted as being a more likely an apparent correlation, since different mechanisms drive the production of either \fe2\ or \h2\ according to the level of nuclear activity. It is not discarded that a combination of X-ray heating from the central source (in AGNs) plus shock heating (radio jet and NLR gas interaction) and/or SNRs can simultaneously drive the emission of \fe2, supporting this correlation in LINERs and SFGs.

\end{itemize}

\section*{Acknowledgements}
We thank an anonymous referee for valuable comments that helped to improve the text. R.R. thanks to FAPERGs (ARD 11/1758-5), CNPq (304796/2011-5) and Instituto Nacional de Ci\^encia e Tecnologia em Astrof\'{i}sica - INCT-A, funded by CNPq and FAPESP. ARA acknowledges CNPq (308877/2009-8) for partial support to this work. I.A. acknowledges support from CAPES/PRODOC and FAPESP (Proc. 2007/04498-2).  OLD is grateful to the FAPESP for support under grant 2009/14787-7. This research has made use of the NASA/IPAC Extragalactic Database (NED) which is operated by the 
Jet Propulsion Laboratory, California Institute of Technology, under contract with the National Aeronautics and Space Administration.

\label{lastpage}

\end{document}